\documentclass[twocolumn]{aastex62}
\pdfoutput=1 
\usepackage[version=4]{mhchem}
\usepackage{amsmath}
\usepackage{multirow}
\usepackage{hyperref}
\usepackage[utf8]{inputenc}
\usepackage{comment}
\usepackage{amsmath}

\newcommand{\revision}[1]{\textcolor{black}{#1}}

\graphicspath{{./}{figures/}}

\received{TBD}
\revised{TBD}
\accepted{TBD}

\shorttitle{The role of clouds on planetary spectra}
\shortauthors{Molaverdikhani et al.}

\begin{document}

\title{The role of clouds on the depletion of methane and water dominance in the transmission spectra of irradiated exoplanets}

\correspondingauthor{Karan Molaverdikhani}
\email{Karan@mpia.de}

\affiliation{Max Planck Institute for Astronomy, Königstuhl 17, 69117 Heidelberg, Germany}

\author{Karan Molaverdikhani}
\affiliation{Max Planck Institute for Astronomy, Königstuhl 17, 69117 Heidelberg, Germany}

\author{Thomas Henning}
\affiliation{Max Planck Institute for Astronomy, Königstuhl 17, 69117 Heidelberg, Germany}

\author{Paul Molli\`ere}
\affiliation{Max Planck Institute for Astronomy, Königstuhl 17, 69117 Heidelberg, Germany}


\begin{abstract}
Observations suggest an abundance of water and paucity of methane in the majority of observed exoplanetary atmospheres. We isolate the effect of atmospheric processes to investigate possible causes. Previously, we studied the effect of effective temperature, surface gravity, metallicity, carbon-to-oxygen ratio, and stellar type assuming cloud-free thermochemical equilibrium and disequilibrium chemistry. However, under these assumptions, methane remains a persisting spectral feature in the transmission spectra of exoplanets over a certain parameter space, the Methane Valley. In this work we investigate the role of clouds on this domain and we find that clouds change the spectral appearance of methane in two direct ways: 1) by heating-up the photosphere of colder planets, and 2) by obscuring molecular features. The presence of clouds also affects methane features indirectly: 1) cloud heating results in more evaporation of condensates and hence releases additional oxygen, causing water dominated spectra of colder carbon-poor exoplanets, and 2) HCN/CO production results in a suppression of depleted methane features by these molecules. The presence of HCN/CO and a lack of methane could be an indication of cloud formation on hot exoplanets. Cloud heating can also deplete ammonia. Therefore, a simultaneous depletion of methane and ammonia is not unique to photochemical processes. We propose that the best targets for methane detection are likely to be massive but smaller planets with a temperature around 1450~K orbiting colder stars. We also construct Spitzer synthetic color-maps and find that clouds can explain some of the high contrast observations by IRAC's channel 1 and 2.
\end{abstract}



\section{Introduction} \label{sec:intro}

After discovery of the first exoplanets in 1990's \citep{wolszczan_planetary_1992} and a fast growing number of discoveries since then, there have been many attempts to observe and characterize their atmospheres \citep[for some recent reviews see e.g.][]{sing_observational_2018,fortney_modeling_2018,helling_exoplanet_2019,madhusudhan_exoplanetary_2019}. Water and methane have been the focus of many investigations due to their relevance to the origin of life and habitability, as well as their major roles \revision{in shaping} the structure of planetary atmospheres \citep[e.g.][]{ackerman_precipitating_2001,birkby_detection_2013,agundez_puzzling_2014,fraine_water_2014,kreidberg_detection_2015,sing_continuum_2016,kreidberg_water_2018,tsiaras_population_2018,alonso-floriano_multiple_2019,benneke_sub-neptune_2019,sanchez-lopez_water_2019}. Abundances retrieved for these species can be also used as a tracer of the carbon-to-oxygen ratio (C/O) and metallicity of these atmospheres; hence potentially linking the formation scenarios with the observations \citep[e.g.][]{raymond_making_2004,bethell_formation_2009,oberg_effects_2011,henning_chemistry_2013,obrien_water_2014,mordasini_imprint_2016,cridland_connecting_2019}. Water's spectral signature has been discovered frequently \citep[e.g.][and references therein]{tsiaras_population_2018} but despite many efforts there has been only one robust detection of methane in irradiated exoplanets so far, and only through high-resolution spectroscopy \citep{guilluy_exoplanet_2019}.

In contrast, methane has been observed on most planets in the solar system. In Earth's atmosphere, living organisms primarily produce it \citep[e.g.][]{Schoell_multiple_1988,catling_biogenic_2001}. On gaseous planets, it is believed to be due to abiogenic processes \citep[e.g.][]{guillot_comparison_1999,glasby_abiogenic_2006,guillot_uranus_2019}. On Mars, its origin is yet to be known \citep[e.g.][]{Krasnopolsky_detection_2004,atreya_methane_2007}. Abiogenic processes \revision{remain} to be tested to explain the possible presence of methane on early Mars \citep[e.g.][]{kite_methane_2019} and methane and oxygen abundances measured by the Curiosity rover over several Martian years \citep[e.g.][]{trainer_seasonal_2020}. These recent measurements have been examined several times to dilute the possibility of human errors. \revision{The question is:} ``How well do we understand \ce{CH4} chemistry to rule out an abiogenic origin on the present Mars?'' While further in-situ measurements, laboratory studies, and model development are crucial to address this question \citep[e.g.][]{hu_hypotheses_2016}, exploring the validity of \ce{CH4} chemistry over a broad range of chemical environments on exoplanets opens a new path. Consequently, consistency of observations and simulations of \ce{CH4} on highly irradiated exoplanets might provide a framework on \ce{CH4} chemistry to be further adapted to study the chemistry of temperate planets.

To investigate the origin of apparent paucity of methane on the majority of observed exoatmospheres, we isolated the effect of atmospheric processes by applying a hierarchical modelling approach. In \citet{molaverdikhani_toward_2019}, we present the results of our extensive thermochemical cloud-free self-consistent simulations with more than 28,000 models. We propose a new classification scheme for irradiated gaseous planets based on their dominant chemistry at their photospheric levels. We focus on planets with an effective temperature between 400\;K and 2600\;K. The classification consists of four classes. Class-I planets (with an effective temperature more than 400\;K and less than 600-1100\;K) are expected to show strong methane and water features in their transmission spectra, \textit{if} they are cloud-free and in thermochemical equilibrium. Class-II planets are hotter than Class-I (T\textsubscript{eff} $<$ 1650\;K). Their transmission spectra are largely sensitive to the atmospheric C/O ratio, effective temperature, and $\beta$-factor (a linear combination of metallicity and surface gravity, $\beta$=log(g)-1.7$\times$[Fe/H], which provides an indication of the photospheric pressure level; also see \citet{molliere_model_2015}). Low C/O ratios usually result in water-dominated spectra and high C/O ratios lead to methane-dominated spectra. \revision{The transition occurs at so-called ``transition C/O ratio'', C/O\textsubscript{tr}.} Class-III planets are hotter than Class-II and free of condensates. Lack of condensates makes the transition C/O ratio insensitive to the $\beta$-factor and temperature, and remains around a value of 0.94, as also found in \citet{molliere_model_2015}. Class-IV planets \revision{require} higher C/O ratios to present \ce{CH4} features in their transmission spectra. This is mainly due to their hotter photosphere and the formation of HCN and CO as the main sinks of carbon instead of \ce{CH4}. We also find a parameter space (800\;K $<$ T\textsubscript{eff}$ < $1500\;K and C/O ratio above a certain threshold value) with a higher chance of methane detection; the Methane Valley. HD\;102195b, i.e. the only known irradiated exoplanet so far with methane in its atmosphere \citep{guilluy_exoplanet_2019}, resides in this parameter space, which supports our prediction. Comparing the rest of observed planets with these results suggest either these planets have C/O ratios lower than their transition C/O ratio (the C/O ratio at which the transition from a water-dominated to methane-dominated atmosphere occurs) or other atmospheric processes \revision{cause the observed methane depletion}. Given the temperature and surface gravity of the characterized exoplanets, the transition C/O ratio of some of them are expected to occur at very low C/O ratios, e.g. less than 0.2.
Such low (sub-stellar) C/O ratios are less in favor from a planetary formation point of view \citep[e.g.][]{oberg_effects_2011, espinoza_metal_2017}. Therefore, it is expected that some of these planets show methane in their transmission spectra, unless other atmospheric processes cause methane depletion.

In \citet{molaverdikhani_fingerprints_2019} we investigate the importance of chemical kinetics in the planetary atmospheres, and in particular on methane depletion, by implementing \revision{the} \texttt{C}hemical \texttt{K}inetic \texttt{M}odel, \texttt{ChemKM}. The results of more than 112,000 chemical kinetic models with full chemical network indicates strong vertical mixing could homogenize vertical abundance of methane; causing quenched abundances. A quenched abundance may result in methane depletion or methane enhancement, depending on its vertical distribution. An example of such case is shown in \citet{molaverdikhani_fingerprints_2019} Figure\;2. Nevertheless, strong vertical mixing makes the boundary between class-II and III planets less profound, but \revision{the Methane Valley continues to occur in a similar part of the parameter space}. Rarity of methane observations brings us to the next step to explore yet another fundamental process in the atmosphere of planets: clouds.

\begin{figure}[t]
\includegraphics[width=\columnwidth]{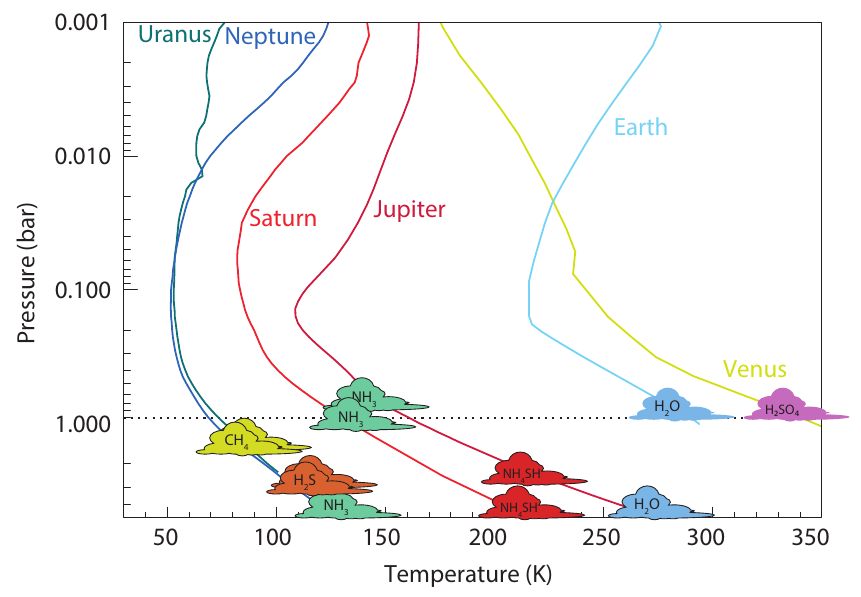}
\caption{Temperature profiles and cloud compositions in the solar system's gaseous planets. Earth's and Venus's profiles are included for reference. The horizontal dashed line marks 1\;bar level where most of the visible cloud decks are located at. This pressure is therefore commonly used as a reference pressure in the solar system community. The temperature profiles are adapted from \citet{robinson_common_2014}.}\label{fig:ch1ssTP}
\end{figure}

All planets in the solar system with a substantial atmosphere possess clouds. Their composition and vertical distribution, however, varies from one planet to another, see Figure~\ref{fig:ch1ssTP}. With temperatures ranging from 50 to 350\;K at around 1\;bar, this includes clouds composed of \ce{CH4}, \ce{H2S}, \ce{NH3}, \ce{NH3SH}, \ce{H2O}, \ce{H2SO4}, etc. Regardless of the temperature structure or the composition of these planets, their visible cloud deck appears to be located at around 1\;bar, which makes it a convenient reference pressure for the solar system community \citep[e.g.][]{seiff_thermal_1998,guillot_comparison_1999}.

As far as the prevalence of clouds is concerned, exoplanets seem to follow a similar story. Transit observations of exoplanets in their near-infrared spectra revealed lower amplitude of water signatures in comparison to the expected amplitudes from cloud-free simulations \citep[e.g.][]{sing_continuum_2016, tsiaras_population_2018, pinhas_h2o_2019}. Water depletion in the protoplanetary disks at the planet’s formation location has been debated as a cause of this lower amplitude
\citep[e.g.][]{seager_dayside_2005,Madhusudhan_toward_2014,Madhusudhan_h2o_2014,sing_continuum_2016}. An alternative possible explanation is the presence of haze and clouds. Their physical properties (such as particle shape, size distribution and chemical composition) affect their optical properties \citep[e.g.][]{heng_understanding_2013}, and their gray or semi-gray opacity in the optical and NIR could obscure atomic and molecular features in this wavelength range. These physical properties depend on the atmospheric structure and dynamics  \citep[e.g.][]{morley_neglected_2012}, and in turn, their radiative feedback could change the temperature structure of the atmosphere and the atmospheric scale height \citep[e.g.][]{morley_neglected_2012}, which also influences the atmospheric composition \citep[e.g.][]{helling_exoplanet_2019,poser_effect_2019}. Consequently, our proposed classification and the Methane Valley might be affected by the formation of clouds in the atmosphere of irradiated exoplanets and might explain the observed paucity of methane.

In what follows, we describe the method and our grid of cloudy models in Section~\ref{sec:ch5method}. In Section~\ref{sec:ch5class}, we present the results of (C/O)$_{\rm tr}$ ratios and how our classification and the Methane Valley are affected by clouds. In Section~\ref{sec:ch5color}, we construct synthetic Spitzer transmission and emission color-diagrams and discuss the consistency of our results with observations. We summarize and conclude our results and findings in Section~\ref{sec:ch5conclusion}.

\section{Methods} \label{sec:ch5method}

In order to investigate the influence of clouds on the atmospheric properties, we have synthesized a population of 37,800 1D self-consistent cloudy planetary atmospheres by using \emph{petitCODE} \citep{molliere_model_2015,molliere_observing_2017}. The model assumes radiative-convective and thermochemical equilibrium and iteratively solves for the atmospheric composition and thermal structure. The code can be used to also calculate the emission and transmission spectra.

\revision{Our calculations use the cloud model described in \citet{ackerman_precipitating_2001}, as implemented and described in \citet{molliere_observing_2017}. In short the clouds are parametrized using three free parameters: the sedimentation factor $f_{sed}$, the atmospheric eddy diffusion coefficient $K_{zz}$, and the width of the log-normal particle size distribution $\sigma_g$. The cloud base is located at the lowest altitude where the saturation vapor pressure curve and the atmospheric temperature structure intersect. The scale height of the cloud is then essentially given by $f_{sed}$ via 
\begin{equation}
d {\rm log} X / d {\rm log}P = f_{\rm sed}, 
\end{equation}
where $X$ is the cloud mass fraction and P the atmospheric pressure. $X$ is equal to zero below the cloud deck. The average particle size is chosen such that the ratio of the mass average of the particle settling and mixing velocities equals $f_{sed}$, where the mixing velocity is determined from $K_{zz}$. $K_{zz}$ is found from a parameterization of the average convective mixing and radiative forcing of the atmosphere, as described in \citet{molliere_observing_2017}. Following \citet{ackerman_precipitating_2001}, we set $\sigma_g$ = 2.}

The chemical inputs of the code are a list of atomic species and a list of reaction products that are the same as the chemical inputs described in \citet{molaverdikhani_toward_2019}\footnote{As a reminder they are the lists of atomic species (H, He, C, N, O, Na, Mg, Al, Si, P, S, Cl, K, Ca, Ti, V, Fe, and Ni) with their mass fractions and reaction products (H, \ce{H2}, He, O, C, N, Mg, Si, Fe, S, Al, Ca, Na, Ni, P, K, Ti, CO, OH, SH, \ce{N2}, \ce{O2}, SiO, TiO, SiS, \ce{H2O}, \ce{C2}, CH, CN, CS, SiC, NH, SiH, NO, SN, SiN, SO, \ce{S2}, \ce{C2H}, HCN, \ce{C2H2}, \ce{CH4}, AlH, AlOH, \ce{Al2O}, CaOH, MgH, MgOH, \ce{PH3}, \ce{CO2}, \ce{TiO2}, \ce{Si2C}, \ce{SiO2}, FeO, \ce{NH2}, \ce{NH3}, \ce{CH2}, \ce{CH3}, \ce{H2S}, VO, \ce{VO2}, NaCl, KCl, \ce{e-}, \ce{H+}, \ce{H-}, \ce{Na+}, \ce{K+}, \ce{PH2}, \ce{P2}, PS, PO, \ce{P4O6}, PH, V, VO(c), VO(L), \ce{MgSiO3(c)}, \ce{Mg2SiO4(c)}, SiC(c), Fe(c), \ce{Al2O3(c)}, \ce{Na2S(c)}, KCl(c), Fe(L), \ce{Mg2SiO4(L)}, SiC(L), \ce{MgSiO3(L)}, \ce{H2O(L)}, \ce{H2O(c)}, TiO(c), TiO(L), \ce{MgAl2O4(c)}, FeO(c), \ce{Fe2O3(c)}, \ce{Fe2SiO4(c)}, \ce{TiO2(c)}, \ce{TiO2(L)}, \ce{H3PO4(c)} and \ce{H3PO4(L)})}.

\begin{figure}
\includegraphics[width=\columnwidth]{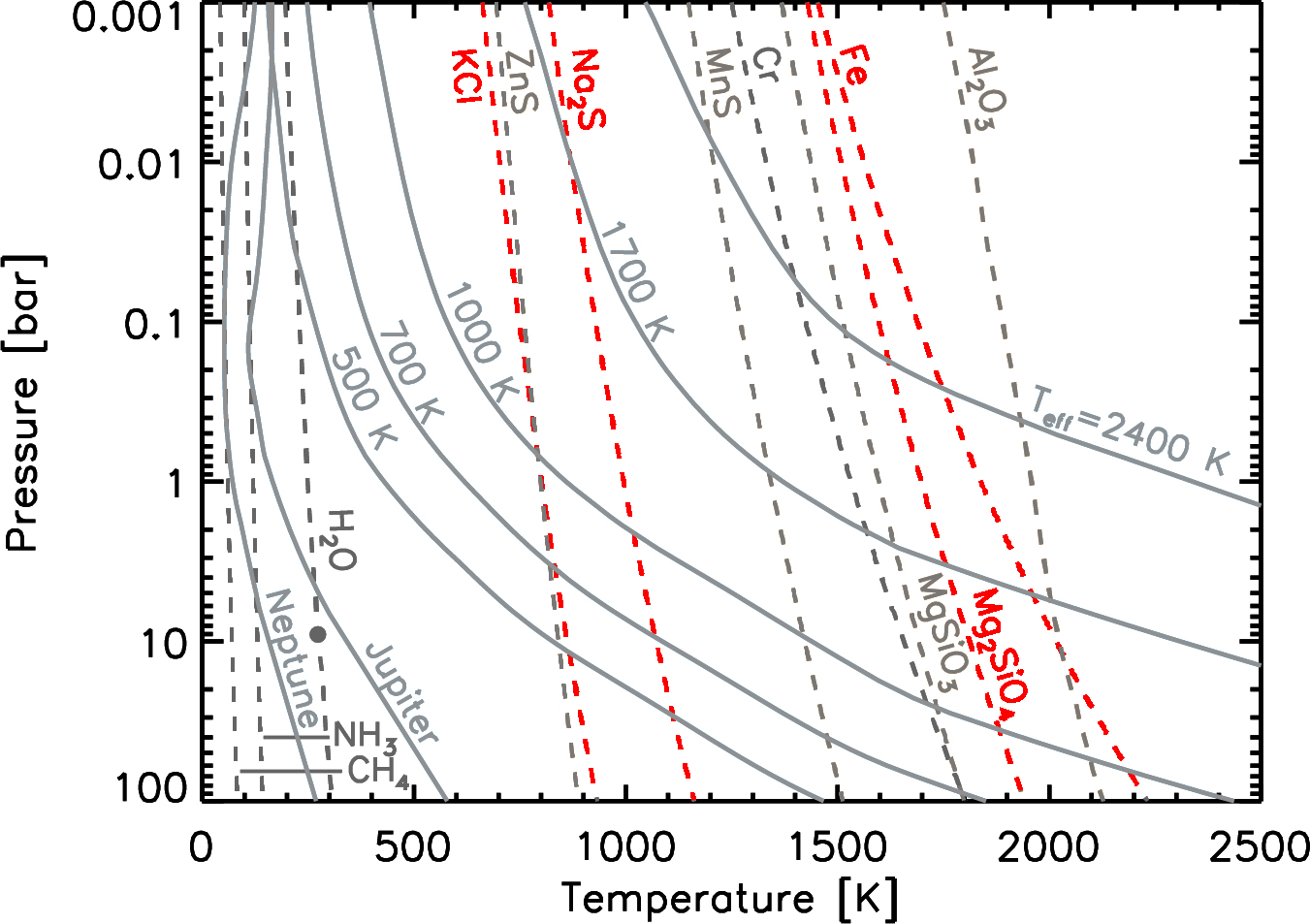}
\caption{Temperature profiles of the planets in solar system and some synthesized TPs for cloud-free exoplanets (solid lines). Condensation curves for variety of chemical substances (dashed lines). The opacities of \ce{Mg2SiO4}, Fe, KCl and \ce{Na2S} are considered in our cloudy grid (red dashed lines), which could affect the temperature profile and vertical composition of the planetary atmospheres. Adapted from \citet{marley_cool_2015}} \label{fig:condensationcurves}
\end{figure}

In addition to the gas opacity species (\ce{CH4}, \ce{H2O}, \ce{CO2}, HCN, CO, \ce{H2}, \ce{H2S}, \ce{NH3}, OH, \ce{C2H2}, \ce{PH3}, Na, K, TiO, and VO) considered in our grid of cloud-free models, cloud opacity species are also provided, which includes \ce{Mg2SiO4}, Fe, KCl and \ce{Na2S}. Figure~\ref{fig:condensationcurves} shows the condensation curves over a broad range of temperatures encountered in the atmosphere of planets and marks the condensation curves of these four condensates as they cross the temperature structure of cloud-free planets. Inclusion of these cloud opacities is expected to introduce a radiative feedback, hence affecting the atmospheric thermal structure and composition. \ce{H2}–\ce{H2} and \ce{H2}–He collision-induced absorption (CIA) is also included in the model. \revision{In theory, the radiative feedback of all condensates must be included in the model. However, we did not include all the condensates as noted. One of the important condensates that we did not consider in this grid of models is MnS. This species has been reported to have noticeable effects on the atmospheric properties over some temperature spans \citep{morley_neglected_2012}, and hence will be considered in the future simulations.}

When interpreting the results of our simulations, it should be noted that the Ackerman-Marley method \citep{ackerman_precipitating_2001}, that has been implemented in petitCODE, assumes one $f_{\rm sed}$ value for all species, which is not necessarily a valid assumption at all conditions \citep[e.g.][]{gao_Sedimentation_2018}. Hence, an overestimation of cloud opacities might be introduced in the case of colder planets. This is discussed in more detail in Sections~\ref{sec:ch5Teff} and ~\ref{sec:ch5conclusion}.

There are two options to treat clouds in petitCODE. One option is to follow the method introduced by \citet{ackerman_precipitating_2001}, for which the following equation must be solved:
\begin{equation}
    -K_{zz}\frac{\partial q_t}{\partial z}-f_{\rm sed}w_*q_c=0
\end{equation}
where $K_{zz}$ is the vertical eddy diffusion coefficient and $w_*$ is the convective velocity scale. \revision{The product of $f_{\rm sed}$ and $w_*$ represents an effective sedimentation velocity for the condensate and can be estimated by $w_*=K_{zz}/H_p$; where $H_p$ is the pressure scale height. More specifically, $f_{\rm sed}$ (the sedimentation factor) is defined as the ratio of the mass-weighted droplet sedimentation velocity, $\langle v_f \rangle$, and $w_*$:}
\begin{equation}
    f_{\rm sed}=\frac{\langle v_f \rangle}{w_*},
\end{equation}
$q_c$ and $q_t$ are the condensate mass fraction and the total mass fraction of the cloud species, \revision{and are related as follow:}
\begin{equation}
    q_t=q_g+q_c
\end{equation}
where $q_g$ is the gas phase mass fraction. \revision{In this model, condensation removes gas from the gas phase chemical model.}

The treatment of vertical mixing \revision{differs from that used in \citet{ackerman_precipitating_2001} and} is described in Appendix A3 of \citet{molliere_observing_2017}. The second option for the treatment of clouds is assuming a constant size for the cloud particles and set the maximum cloud mass fraction \citep[see][]{molliere_observing_2017}. We choose the first approach, which is more physically \revision{motivated}.

When interpreting the results of our simulations, it should be noted that petitCODE, assumes one $f_{\rm sed}$ value for all species, which is not necessarily a valid assumption at all conditions \citep[e.g.][]{gao_Sedimentation_2018}. Hence, an overestimation of cloud opacities might be introduced in the case of colder planets. This is discussed in more detail in Sections~\ref{sec:ch5Teff} and ~\ref{sec:ch5conclusion}.

\subsection{Grid properties} \label{subsec:ch5grid}

We consider the effective temperature (T\textsubscript{eff}), surface gravity (log(g)), metallicity ([Fe/H]), carbon-to-oxygen-ratio (C/O), stellar type, and the sedimentation factor ($f_{\rm sed}$) as the free parameters of this grid of models. These parameters and their values are discussed below.

\subsubsection{Effective temperature (T\textsubscript{eff})}\label{sec:ch5Teff} 
For the cloud-free \citep[][]{molaverdikhani_toward_2019} and disequilibrium \citep[][]{molaverdikhani_fingerprints_2019} grids, we set the interior temperature at 200\;K to be consistent with the \citet{fortney_effect_2005} and \citet{molliere_model_2015} simulations. In the cloudy grid (this work) we assume the same interior temperature, consistent with the previous grids.

In both previous grids (cloud-free and disequilibrium), the range of temperature was selected to span from relatively cold planets, 400\;K, to ultra hot planets, 2600\;K. Therefore, the best choice of temperature range for the cloudy grid would be using the same range. However, as shown in Figure~\ref{fig:condensationcurves}, none of the considered cloud opacities (\ce{Mg2SiO4}, Fe, KCl and \ce{Na2S}) would result in cloud formation at temperatures above 2400\;K under a 1D setup. Thus, we limit the upper effective temperature of the planets to 2400\;K. As shown in Figure~\ref{fig:condensationcurves}, at temperatures below 800\;K, all planets are expected to be covered by all four considered cloud types (\ce{Mg2SiO4}, Fe, KCl and \ce{Na2S}) in 1D setup. Therefore, any planet with an effective temperature colder than 800\;K would result in a spectrum with strongly muted atomic and molecular features. Hence, less information regarding the composition of atmosphere at the photospheric level can be interpreted from the planetary spectra of these planets. Therefore, we limit the lowest effective temperature to 800\;K. We choose an increment of 200\;K to be consistent with our previous grids.

\subsubsection{Surface gravity (log(g))}
In the previous grids, surface gravity in the models spans over a very wide range of possibilities, from 2.0 to 5.0. But planets with a surface gravity of 2.0 or 5.0 are very exceptional \citep[see Figure 1 in ][]{molaverdikhani_toward_2019}. Since the computational time of cloudy models is longer than that of cloud-free models, we opt for excluding the rarely observable parameter space from the cloudy grid. We thus explore this parameter from 2.5 to 4.5 with an increment of
0.5; similar increment to that of previous grids.

\subsubsection{Metallicity {\normalfont([Fe/H])}}
In the previous grids, we chose to explore a wide range of metallicities from sub-solar, [Fe/H]=-1.0, to super-solar [Fe/H]=2.0 with increment of 0.5. We choose the same setup for the range and increment of metallicity in the cloudy grid.

\subsubsection{Carbon-to-oxygen-ratio {\normalfont(C/O)}}\label{sec:ch5CO} 
As discussed in the introduction, one major goal of performing the cloudy grid is to find the transition C/O ratios; i.e. a boundary in the parameter space where the planetary atmosphere transitions from a water-dominated to methane-dominated one as C/O ratio increases. For this reason, in our cloud-free and disequilibrium grids, we selected irregular parameter steps spanning from 0.25 to 1.25 with smaller steps around unity: C/O=[0.25, 0.5, 0.7, 0.75, 0.80, 0.85, 0.90, 0.95, 1.0, 1.05, 1.10, 1.25]. These smaller steps around unity were in order to capture the traditionally believed boundary at around the unity. However, we showed that this transition could occur at any C/O \citep{molaverdikhani_toward_2019,molaverdikhani_fingerprints_2019}. Hence a finer step-size at unity is no longer needed. As a result we choose to vary the C/O from 0.2 to 1.2 with an increment of 0.2 everywhere.

\subsubsection{Stellar type}
We choose the same stellar types in the cloudy grid as the previous grids, i.e. M5, K5, G5 and F5, in order to cover a wide range of stellar types.

\subsubsection{sedimentation factor ($f_{\rm sed}$)}
As noted before, we use the cloud implementation in the petitCODE that is introduced by \citep{ackerman_precipitating_2001}. The product $f_{\rm sed}w_*$ represents the mass-average sedimentation velocity for the condensate, which \revision{causes} an offset to the turbulent mixing. In the case of \revision{zero} sedimentation velocity (with $f_{\rm sed}$=0), condensates remain ``frozen-in'' and move with the gas particles; i.e. following the same velocity. As $f_{\rm sed}$ increases, and so the sedimentation velocity, larger cloud particles, lower cloud mass fraction and larger settled clouds (i.e. less extended) are expected.

Estimated and assumed values of $f_{\rm sed}$ depend on a variety of parameters such as the cloud type and the atmospheric environment. But as a rule of thumb it can vary between 0.01 and 10; although often it retains a value less than 1.0 \citep{lunine_effect_1989,ackerman_precipitating_2001,molliere_observing_2017}. Therefore, we investigate the effect of the sedimentation factor from 10$^{-1.5}$ ($\sim$0.03) to 10$^{0.5}$ ($\sim$3) with a logarithmic increment of 10$^{0.5}$.



\subsection{Spectral Decomposition}\label{sec:ch5decomp}
\begin{figure}[t]
\includegraphics[width=\columnwidth]{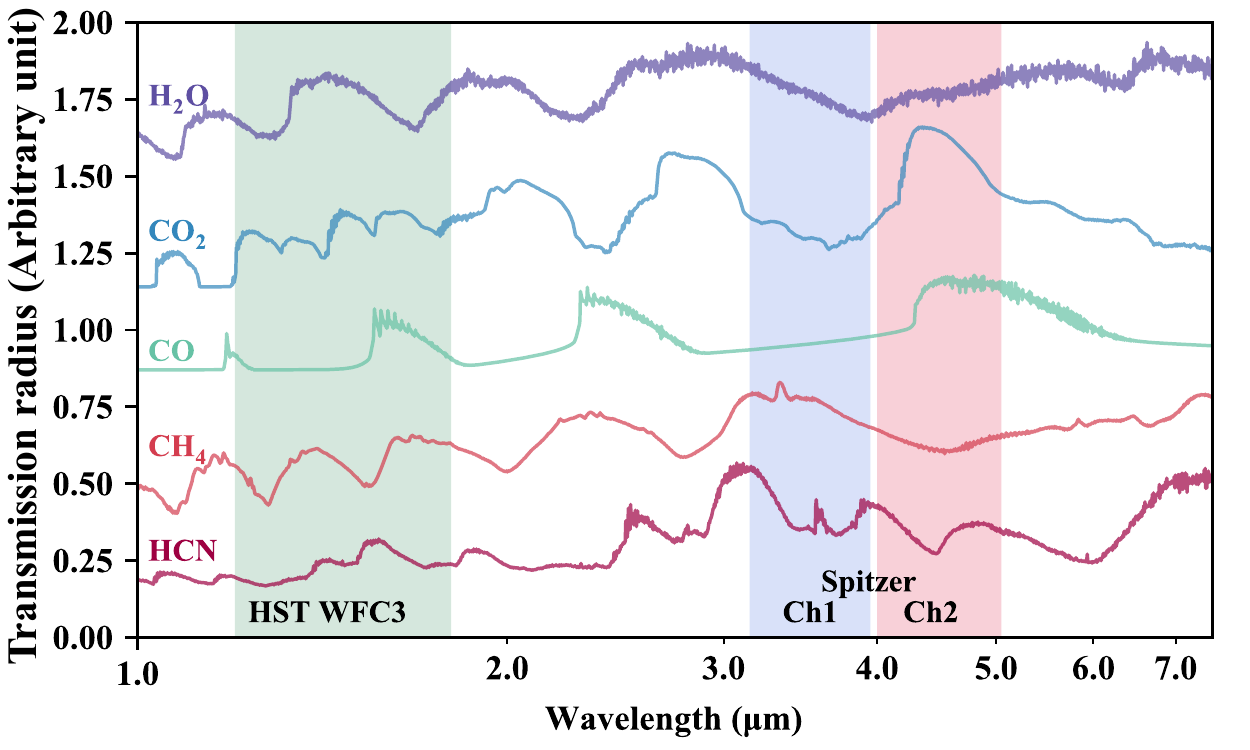}
\caption{Templates of major opacity sources used in the spectral decomposition technique, see Section~\ref{sec:ch5decomp}. The wavelength coverage of HST-WFC3 and Spitzer-IRAC's channel 1 and 2 are shown for reference.} \label{fig:ch5spectemp}
\end{figure}

In order to quantitatively estimate the transition C/O ratios, the contribution of each atmospheric constituent in the spectra should be estimated. To perform this, we follow the same as in \citep{molaverdikhani_toward_2019}, using the spectral decomposition technique. Constructed templates of transmission spectra of individual species, assuming an isothermal TP \revision{profile} with T=1600\;K, are shown for some major opacity sources in Figure~\ref{fig:ch5spectemp}. \revision{We find the combination of these templates that best fits the model, and the fit coefficients are taken to be the contribution from each species.} By definition, the ratio of contribution coefficient of \revision{methane ($c_{\rm CH_4}$) to water ($c_{\rm H_2O}$), i.e. $c_{\rm CH_4}/c_{\rm H_2O}$,} is unity at any transition C/O ratio. For a methane-dominated spectrum $c_{\rm CH_4}/c_{\rm H_2O}>1$ and for a water-dominated spectrum it is $c_{\rm CH_4}/c_{\rm H_2O}<1$.

\section{Classification scheme and the Methane Valley}\label{sec:ch5class}

\subsection{The role of clouds}
The complexity of the radiative feedback of clouds in self-consistent calculations makes studying their effects in an automated way a challenge. We therefore begin our investigation by exploring these effects through the analysis of several examples. We then generalize our findings by using an automated and quantitative approach in Section~\ref{sec:ch5COratios}.

\begin{figure*}
\gridline{\fig{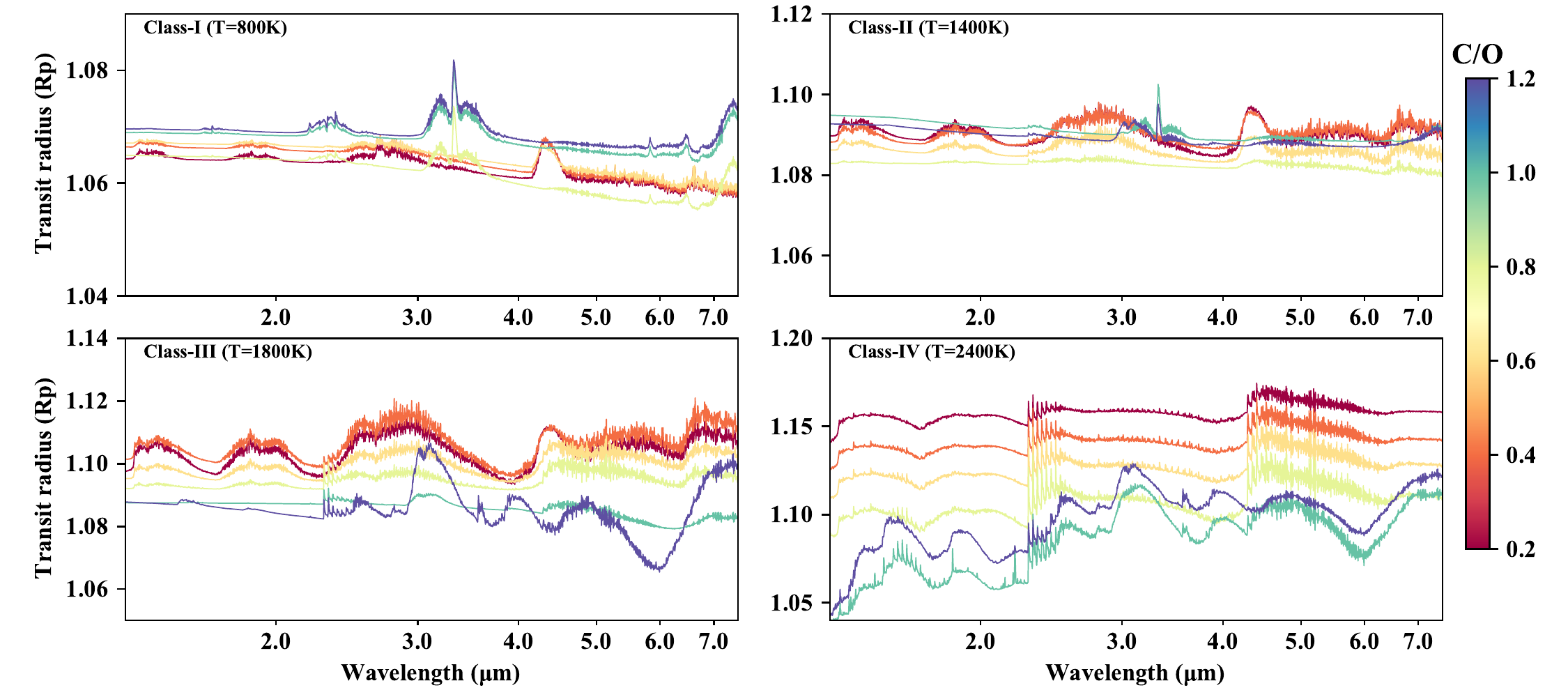}{\textwidth}{(a) Cloudy models (this work)}}
\gridline{\fig{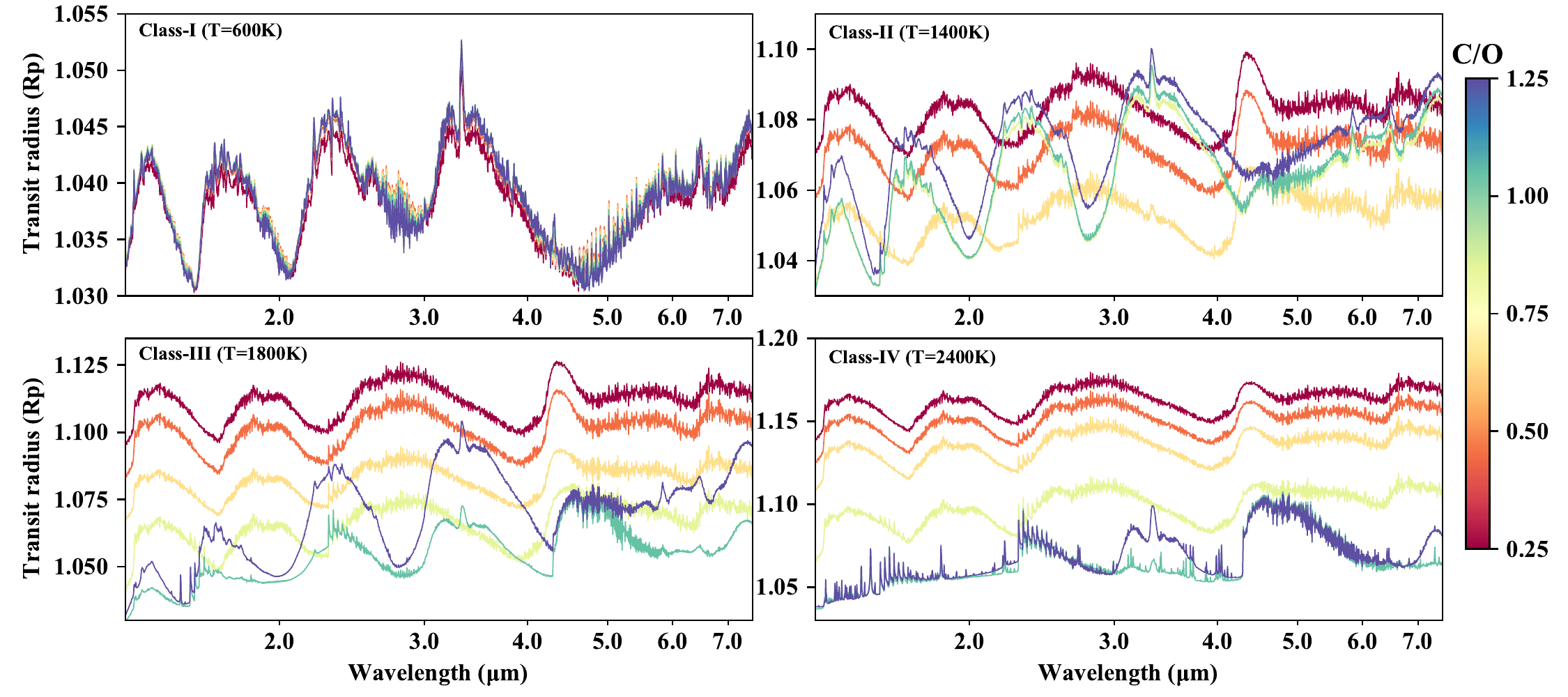}{\textwidth}{(b) Similar models to (a) but with cloud free assumption, figure from \citet{molaverdikhani_toward_2019}. Temperature of the coldest Class-I model in the two grids are slightly different; see Section~\ref{subsec:ch5grid} for details.}}
\caption{Examples of transmission spectra: (a) Cloudy models at $f_{\rm sed}\sim$0.03. Carbon-rich class-I and class-II planets show spectral signatures of \ce{CH4} but in hotter classes (III and IV) \ce{CH4} is replaced with CO and HCN in comparison to the cloud-free atmospheres (b).
\label{fig:examplesf003}}
\end{figure*}

Figure~\ref{fig:examplesf003}(a) illustrates how the transmission spectrum of cloudy models changes with carbon-to-oxygen ratio in four classes of planets. We assume log(g)=3.0, [Fe/H]=1.0, G5 star, and $f_{\rm sed}\sim$0.03 for these models. This surface gravity and metallicity are chosen to be representative of atmospheric compositions of sub-Jovian gaseous planets in the solar system\footnote{Saturn, Uranus and Neptune, all have a surface gravity of around 3.0 with super-solar metallicities.}.

Comparing these spectra with the cloud-free spectra, see Figure~\ref{fig:examplesf003}(b), reveals several striking differences. Due to low sedimentation factor in these examples, clouds remain high in the atmosphere (at low pressures) and prevent molecular features to appear in full. Unsurprisingly, colder planets are influenced by the cloud opacities the most. This is evident in the transmission spectra of Class-I planets, as they are the coldest planets in our grid of models. Low C/O ratios (redder colors) are expected to result in the presence of both methane and water in the transmission spectra, assuming cloud-free atmospheres \citep[e.g.][]{molliere_model_2015,molaverdikhani_toward_2019}. In cloudy models, however, methane vanishes at low C/O ratios. The level of cloud continuum extinction in these examples suggests that methane is depleted at lower C/O ratios rather than being obscured by clouds.

\begin{figure*}
\centering
\includegraphics[width=0.8\textwidth]{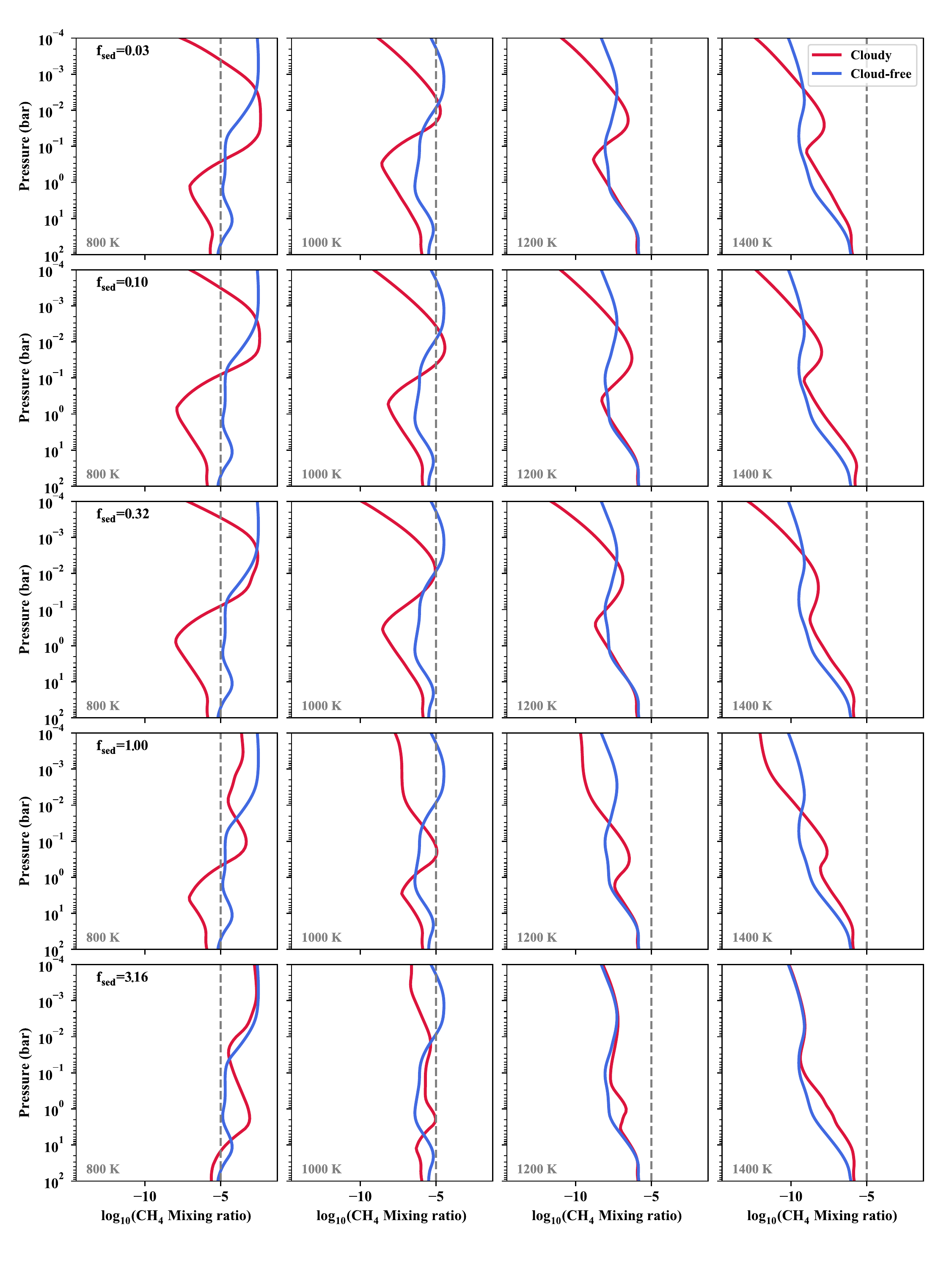}
\caption{Comparing methane abundances \revision{in a} low C/O ratio environment (C/O=0.25), in cloudy models and in cloud-free models. In all examples, metallicity, C/O ratio, and stellar type are [Fe/H]=1.0, C/O=0.25, and G5. Methane abundances are lower at most altitudes and in most cases, although there are some regions with more \ce{CH4} production due to local atmospheric cooling. Hotter planets, in general, produce much less methane as expected; the vertical dashed line marks a 10$^{-5}$ mixing ratio for reference. Abundances are shown for different values of sedimentation factor, $f_{\rm sed}$. \label{fig:ch5abundCH4}}
\end{figure*}

\begin{figure*}
\centering
\includegraphics[width=0.8\textwidth]{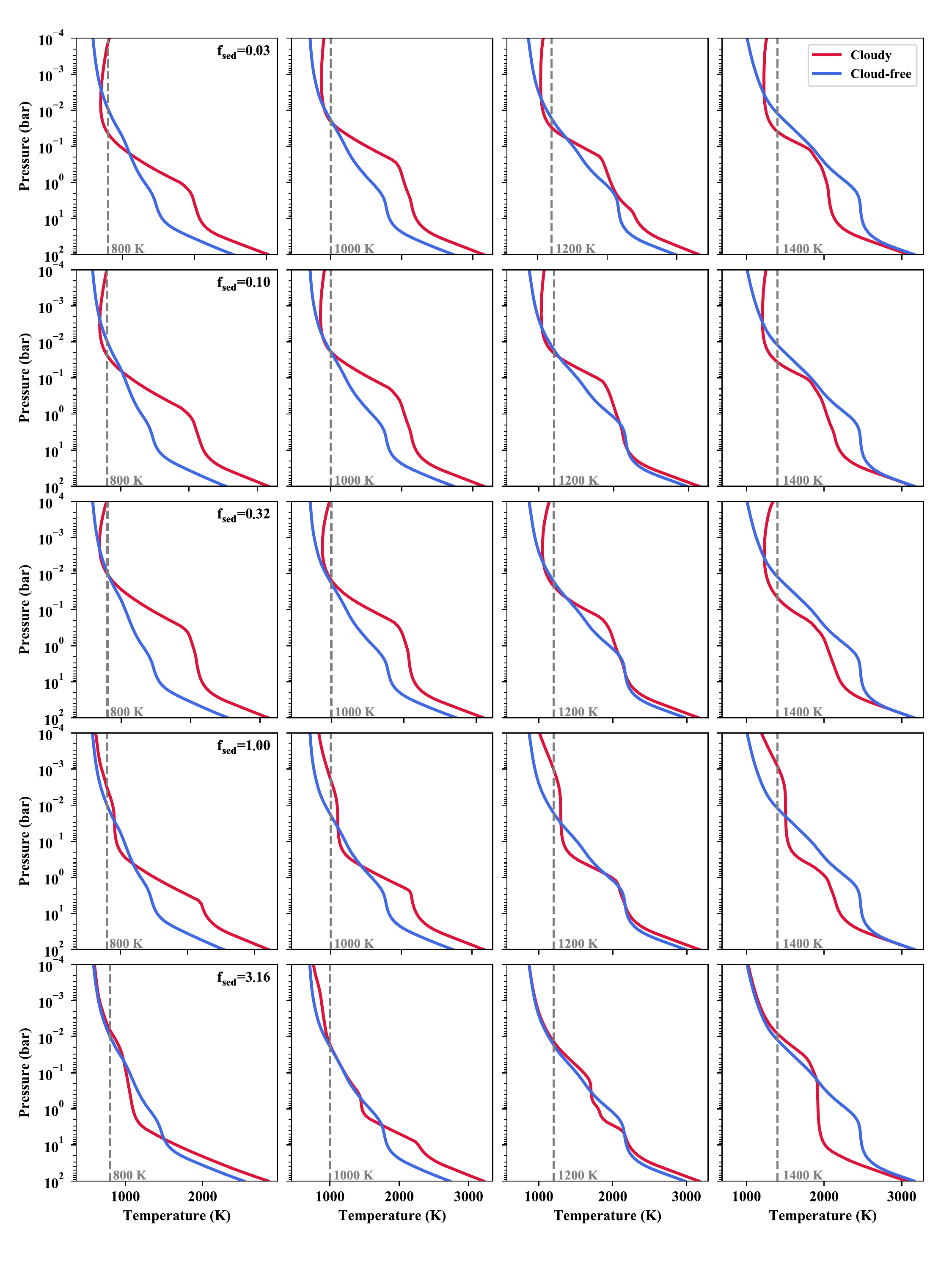}
\caption{Corresponding temperature structures to Figure~\ref{fig:ch5abundCH4} models; Metallicity, C/O ratio, and stellar type are [Fe/H]=1.0, C/O=0.25, and G5.
Atmospheric temperatures are relatively hotter for all planets with temperatures relevant to the formation of \ce{CH4}, e.g. T\textsubscript{eff} $<$ 1200\;K, although the cloud feedback is highly non-linear. Temperature profiles are shown for different values of sedimentation factor, $f_{\rm sed}$.} \label{fig:ch5TPs}
\end{figure*}

To examine this, abundances of methane \revision{in a} low C/O ratio environment are compared between cloudy and cloud-free models in Figure~\ref{fig:ch5abundCH4}. Metallicity, C/O ratio, and stellar type are assumed to be 1.0, 0.25 and G5, respectively. For the cloudy models the results of C/O=0.2 and C/O=0.4 are linearly combined to estimate the abundance profiles at C/O=0.25. These models show that for Class-I planets, at most altitudes methane abundances are diminished at the presence of clouds. Although there are some regions with more \ce{CH4} production in the cloudy models. These local depletion and production of methane are due to local atmospheric heating and cooling as show in Figure~\ref{fig:ch5TPs}. This mechanism is also consistent with the observation of depleted methane on self-luminous objects \citep{barman_young_2011,currie_combined_2011,marley_masses_2012}. The excess heating partially releases the sequestered oxygen from \revision{some of} the condensates \revision{such as VO(c), VO(L), \ce{MgSiO3(c)}, \ce{Mg2SiO4(c)}, \ce{Al2O3(c)}, \ce{Mg2SiO4(L)}, \ce{MgSiO3(L)}, \ce{H2O(L)}, \ce{H2O(c)}, TiO(c), TiO(L), \ce{MgAl2O4(c)}, FeO(c), \ce{Fe2O3(c)}, \ce{Fe2SiO4(c)}, \ce{TiO2(c)}, \ce{TiO2(L)}, \ce{H3PO4(c)} and \ce{H3PO4(L)}). The excess oxygen} accelerates the formation of water instead of methane. This additional \ce{H2O} manifests itself in the transmission spectra of cold planets with low C/O ratios, see e.g. Figure~\ref{fig:examplesf003}(a). In addition to \ce{H2O}, \ce{CO2} spectral feature at around 4.5~$\mu$m is also noticeable that was not being present in cloud-free models, see Figure~\ref{fig:examplesf003}(b). Therefore the following reaction is a possible chemical pathway for this conversion:
\begin{equation} \ce{CH4 + 2O2 -> CO2 + 2H2O \label{eq:ch5CH4H2O}}.  \end{equation}

A hotter deeper atmosphere due to clouds' radiative feedback could result in the depletion of \ce{NH3} too. This is consistent with the observations of GJ\;3470\;b, a warm sub-Neptune \citep{benneke_sub-neptune_2019}. Therefore, an atmosphere depleted of \ce{CH4} and \ce{NH3} could be an indication of heating at deeper regions by a thick layer of silicate, alkali sulfates, salts, or other non-soot clouds rather than depletion by photochemical processes. Constraining the composition of clouds could aid to link the observable parts to the deeper levels and address which process is responsible.

Figure~\ref{fig:ch5TPs} shows the corresponding temperature structures of Figure~\ref{fig:ch5abundCH4} models, where the temperature profiles appear to be relatively hotter for all models with temperatures relevant to the formation of \ce{CH4}, e.g. T\textsubscript{eff} $<$ 1200\;K. At hotter temperatures, where \ce{Na2S} and KCl clouds are thinning, the deeper parts of the atmosphere could be cooled down by the presence of clouds, e.g. models with T\textsubscript{eff} $=$ 1400\;K. Nevertheless, clouds' feedbacks are found to be highly non-linear throughout the entire parameter space. The effects of different values of sedimentation factor, as shown in Figures~\ref{fig:ch5abundCH4} and \ref{fig:ch5TPs}, will be discussed later.

\revision{In} the Class-I planets' \revision{spectra}, Figure~\ref{fig:examplesf003}(a), even the strongest water features between 2.5 to 3.5\;$\mu$m have been heavily muted. \ce{CO2}, on the other hand, has an appreciable fingerprint between 4.0 to 5.0\;$\mu$m due to weaker other opacities (such as water and clouds) at these wavelengths and excess production through Reaction~\ref{eq:ch5CH4H2O}. At higher C/O ratios (bluer colors) \ce{CH4} comes into sight; although it is significantly muted by clouds. The most prominent spectral signature of methane at 3.3\;$\mu$m seems to be persisting and a good target-wavelength for the future observations of possible carbon-rich but cloudy planets, such as GJ\;1214b-like or GJ\;436b-like planets \citep[e.g.][]{kawashima_detectable_2019}.

\revision{The boundary between Class-I and Class-II planets is the evaporation of condensates in the photosphere of Class-II planets \citep{molaverdikhani_toward_2019}.} This partial evaporation of condensates at hotter temperatures diminishes the contribution of clouds in the opacity budget. Thus, molecular features are expected to be more pronounced with respect to Class-I cloudy planets. While water features follow this prediction at lower C/O ratios and become more pronounced, the methane signal at higher C/O ratios seems to be even more muted than in Class-I; e.g. compare the weak methane feature between 2.0 and 3.0\;$\mu$m in these two classes. This is yet another hint for the depletion of methane at the photospheric levels, rather than obscuration by clouds.

Class-III planets, with effective temperatures between 1650\;K and 2200\;K, show a similar pattern: more pronounced water features at low C/O ratios and no sign of methane at higher C/O ratios. At these temperatures, KCl and \ce{Na2S} are completely evaporated but \ce{Mg2SiO4} and Fe are partially in their solid phase. Partial evaporation of condensates (including the ones that are not considered as the cloud opacities in this study) releases oxygen and causes water production. These diminished cloud opacities and enhanced water production explain stronger water features at low C/O ratios of Class-III planets. This weaker cloud absorption makes the atmospheric cooling to be more efficient, which is consistent with the cooled regions of hotter planets (e.g. at 1400~K) shown in Figure~\ref{fig:ch5TPs}.

On the other hand, higher C/O ratios result in lower water abundances. Water opacity contribution diminishes to a point that no strong molecular feature remains in the spectrum. Therefore, cloud continuum becomes the dominant opacity source at such transition C/O ratio. \revision{This is similar to the results by \citet{molliere_model_2015}, who reported that a C/O close to the transition from O- to C-dominated gas chemistry (around C/O $\sim$0.7-0.9) leads to a minimum in molecular absorber opacities.} In this example, i.e. Class-III planets in Figure~\ref{fig:examplesf003}, CO and HCN become the dominant carbon-bearing species at high C/O ratios and methane's presence is obscured by these species in a non-gray way. \citet{noll_onset_2000} reported a similar mechanism on L Dwarfs for which methane can be obscured by other opacities such as clouds, molecular hydrogen continuum or line opacities. The replacement of \ce{CH4} with CO and HCN in comparison with the cloud-free spectra \citep[e.g. see Figure 4 in][]{molaverdikhani_toward_2019}, and lower cloud opacity contributions support the methane depletion as the main cause of methane's paucity on hot exoplanets.

When the sedimentation factor is extremely low, e.g. $f_{\rm sed}$=0.03, clouds are extended to the very low pressures and contribute significantly to the radiative structure of the atmosphere. Thus the remaining methane features could be also muted; see the void of methane feature at 3.3\;$\mu$m for high C/O ratio Class-III spectra, e.g., in Figure~\ref{fig:examplesf003} ( see also Figures~\ref{fig:examplesf010}, \ref{fig:examplesf030} and \ref{fig:examplesf100} in the Appendix, to compare cases with other sedimentation factors). But occasionally and under specific circumstances the 3.3\;$\mu$m feature emerges debilitated in the models with low sedimentation factors; see e.g. Figure~\ref{fig:examplesf300}.

Transmission spectra of Class IV planets, Figure~\ref{fig:examplesf003}, are conspicuously different from that of Class-III. In particular, \ce{H2O} and \ce{CO2} are replaced by CO at low C/O ratios and there is no sign of cloud continuum even at higher C/O ratios. Methane remains depleted because these temperatures, T\textsubscript{eff} $>$ 2200\;K, make an inauspicious environment for methane production.

\begin{figure*}
\centering
\includegraphics[width=0.8\textwidth]{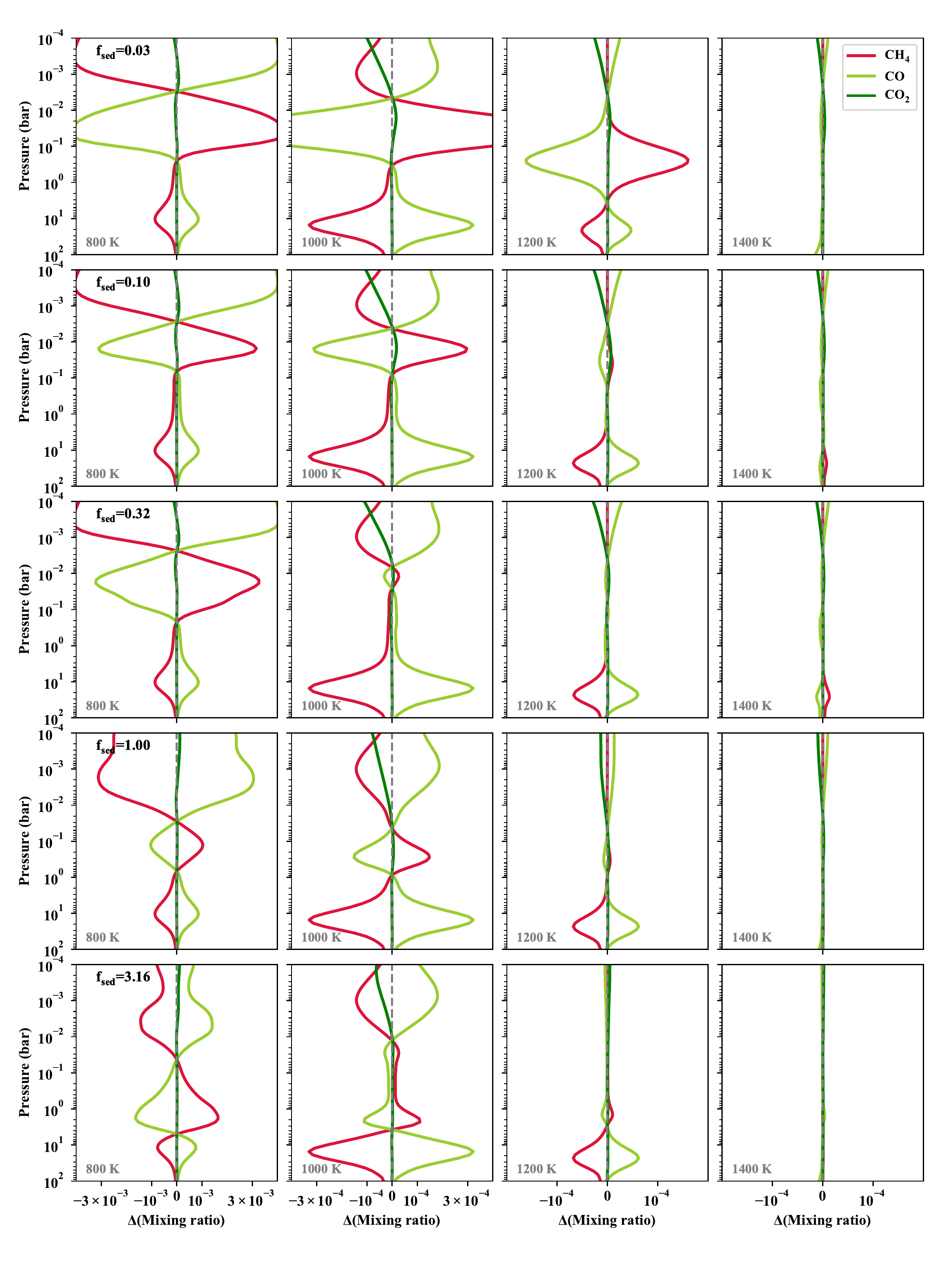}
\caption{Abundance difference between cloudy and cloud-free models for \ce{CH4}, \ce{CO} and \ce{CO2}. Models are the same as in Figure~\ref{fig:ch5abundCH4} and \ref{fig:ch5TPs}, except C/O ratio is selected to be 0.5 for a higher chemical contribution of methane in the composition of atmosphere, for illustration purposes. Vertical dashed lines show zero abundance variation between the two models. Carbon in \ce{CH4} is mostly deposited in CO at the hotter regions of colder planets. On hotter planets, \ce{CH4} plays a negligible role in the carbon chemistry; unless the C/O ratio is very high (not shown here).} \label{fig:ch5dabund}
\end{figure*}

Figure~\ref{fig:ch5dabund} illustrates the main \ce{CH4}-CO conversion pathway for the similar models in Figures~\ref{fig:ch5abundCH4} and \ref{fig:ch5TPs}; except C/O ratio is selected to be 0.5 (instead of 0.25) to achieve higher chemical contribution of methane in the composition of atmosphere for illustration purposes. These conversions can be understood by following this reaction:
\begin{equation} \ce{CH4 + H2O <=>T[$\gtrsim 1000K$][$ \lesssim 1000K$] 3H2 + CO \label{eq:ch5COCH4}}  \end{equation}

At deeper regions, where atmosphere becomes hotter due to the presence of thick clouds, methane depletes and CO captures the freed carbon. This continues at lower pressures, but \ce{CO2} also contributes and becomes CO. These all hint for a prominence of CO features in the transmission spectra of irradiated planets.

Mid to late T dwarfs are also expected to have methane-rich atmospheres, \revision{assuming they are in a chemical equilibrium.} However, observations of CO with relatively high mixing ratios suggested that their atmospheric compositions are likely out of equilibrium \citep{noll_models_1974,saumon_molecular_2000,saumon_ammonia_2006,geballe_spectroscopic_2009,barman_young_2011}. This interpretation is plausible if abundance enhancement at larger pressures is monotonic. If not, then vertical mixing could lead to depleted, enhanced, or invariant \ce{CH4} abundances, depending on the shape of the abundance profile and the strength of the atmospheric mixing \citep[e.g.][]{molaverdikhani_fingerprints_2019}. Therefore, an alternative explanation could be the presence of clouds where higher temperatures at high pressure result in the formation of CO and methane depletion without a need for a strong vertical quenching. A combination of both processes to explain the observed CO on the mid to late T dwarfs seems to be more plausible.

\subsection{The role of sedimentation factor, $f_{\rm sed}$}
\begin{figure*}
\includegraphics[width=\textwidth]{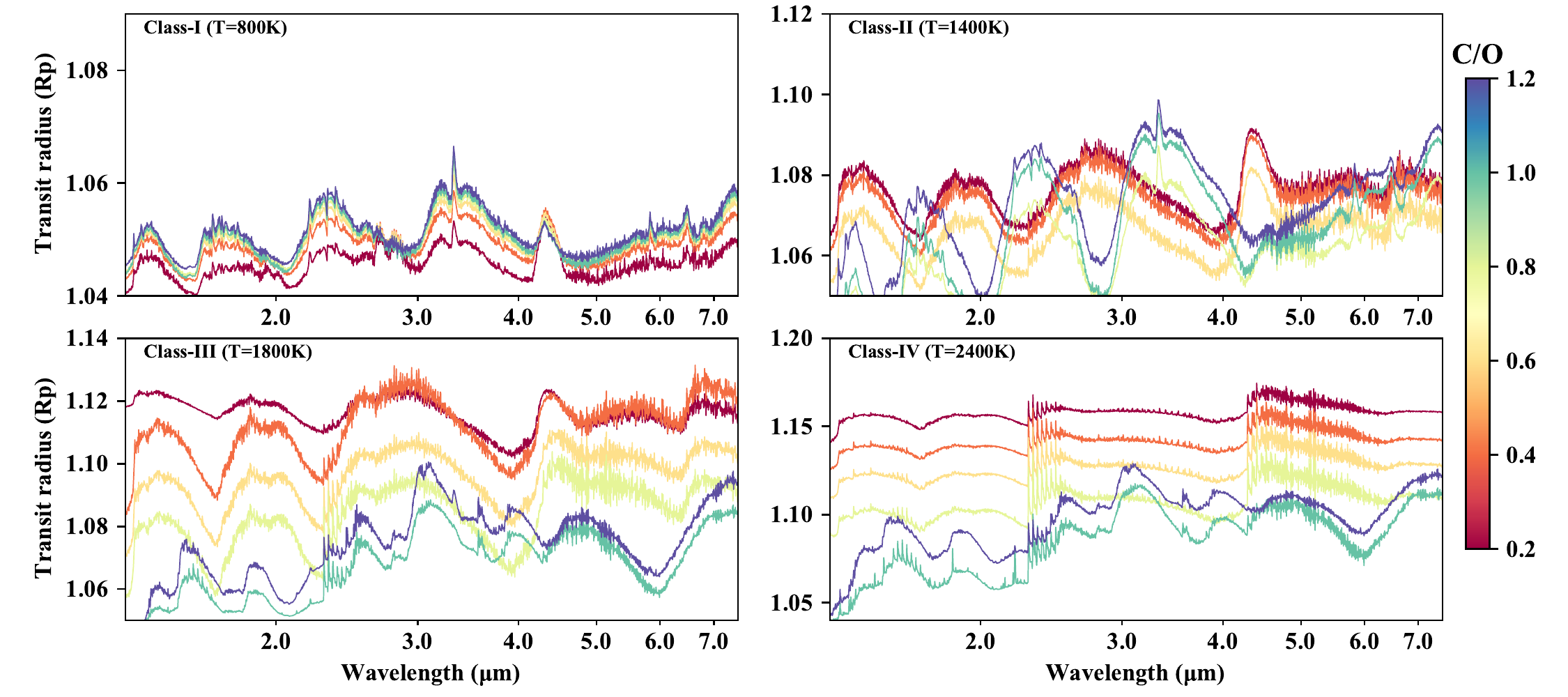}
\caption{Examples of transmission spectra of cloudy models at $f_{\rm sed}=$3.16. High sedimentation factor removes the clouds from the low pressure regime; hence spectra appear similar to their cloud-free counterparts to some degree.} \label{fig:examplesf300}
\end{figure*}

Figure~\ref{fig:ch5TPs} presents the temperature structure of Class-I and II cloudy models (red lines) at different sedimentation factors, ranging from 0.03 to just above 3. In general, a higher $f_{\rm sed}$ results in the cloud vertical \revision{extent} to be narrower and remain at deeper levels \citep{ackerman_precipitating_2001}. This makes the effect of clouds at lower pressures less significant. Consistency of cloud-free and cloudy temperature structures at low pressures ($<$10\;mbar) for the cases with high sedimentation factor, e.g. $f_{\rm sed}$=3.16, supports this statement. Simultaneously, temperature differences at deeper parts of the atmospheres also decrease at high $f_{\rm sed}$ values. Thus, cloudy models with very high $f_{\rm sed}$ are expected to resemble cloud-free models in general. The highest sedimentation factor in our grid of model is 3.16 and Figure~\ref{fig:examplesf300} illustrates corresponding transmission spectra to this value. The rest of model parameters are the same as in Figure~\ref{fig:examplesf003}. The contribution of cloud opacities is barely noticeable in these spectra (Figure~\ref{fig:examplesf300} with $f_{\rm sed}$=3.16) and Class-I and II spectra resemble the spectra of cloud-free models. Methane is the prominent feature in Class-I at all C/O ratios. The \ce{CO2} spectral feature at 4.5\;$\mu$m can be seen as well for this class that is not present in its cloud-free counterpart, e.g. compare Class-I spectra in Figure~\ref{fig:examplesf300} and Figure~\ref{fig:examplesf003}(a). Methane in Class-II planets appears in C/O ratios above the transition C/O ratio but not at low C/O ratios. HCN have a persisting appearance in Class-III and IV planets; hinting for a chemical fingerprint of clouds even at high $f_{\rm sed}$ values.

However, as noted, the radiative response to the cloud opacity is highly non-linear and clouds' feedback depends on a variety of parameters, including the sedimentation factor. The response of temperature structure to the variation of $f_{\rm sed}$ is shown in Figures~\ref{fig:ch5TPs} for a set of models. Looking at the examples with an effective temperature of 800\;K, an increase in $f_{\rm sed}$ from 0.03 to 0.32 heats up the atmosphere in some regions but cools elsewhere. For instance, three atmospheric regions are notable in these particular cases: above a few mbar, below a few hundred mbar, and pressures levels in between. This translates to the depletion of methane at some levels but not the others, as shown in Figure~\ref{fig:ch5dabund}. Associated pressures to these regions, however, are not universal and vary strongly with atmospheric parameters. See, for example, how the temperature profiles vary in hotter cases in Figure~\ref{fig:ch5TPs}.

\begin{figure*}
\includegraphics[width=\textwidth]{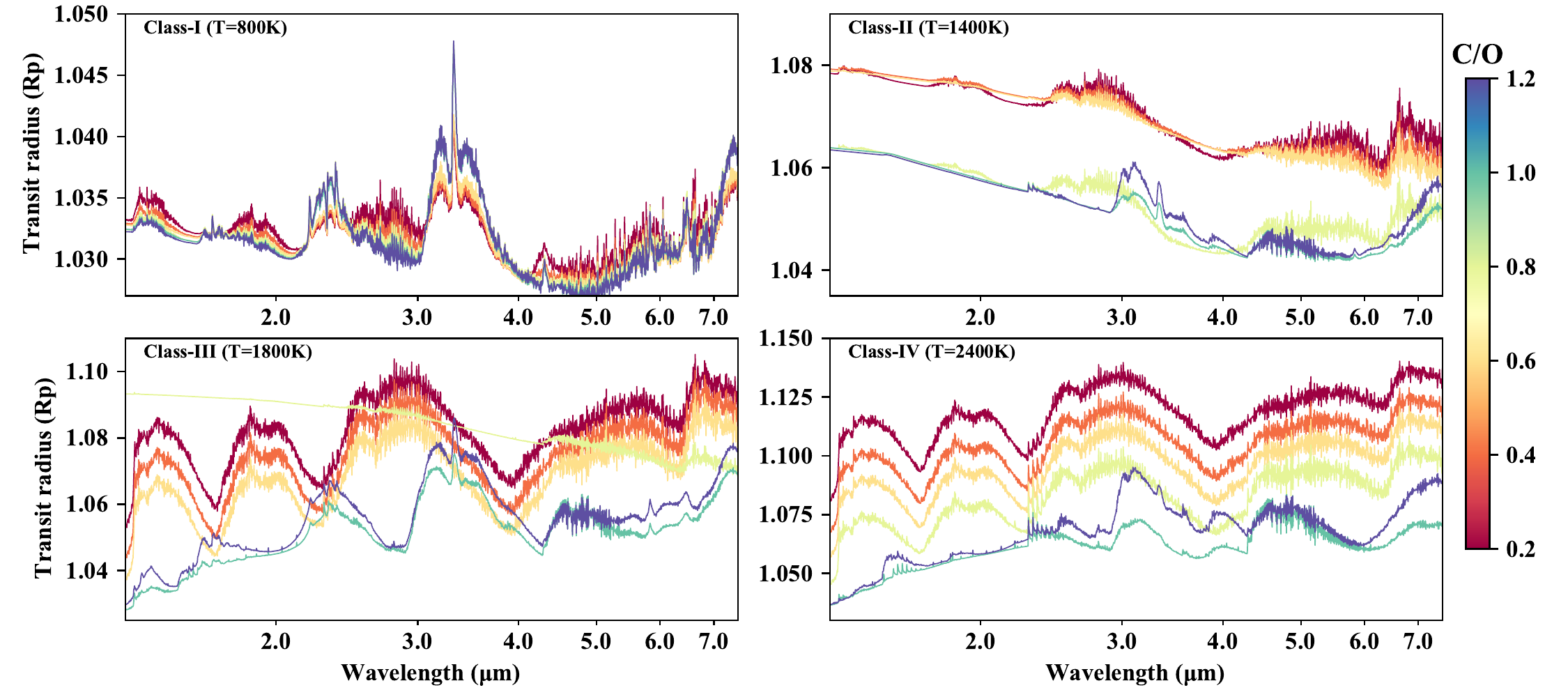}
\caption{Examples of transmission spectra of cloudy models at $f_{\rm sed}\sim$0.03 and [Fe/H]=-1.0. Due to lower $\beta$-factor, and hence the placement of the photosphere at lower pressures, \ce{CH4} is more efficiently produced with respect to HCN.} \label{fig:examplesf003fehn1}
\end{figure*}

\begin{figure*}
\includegraphics[width=\textwidth]{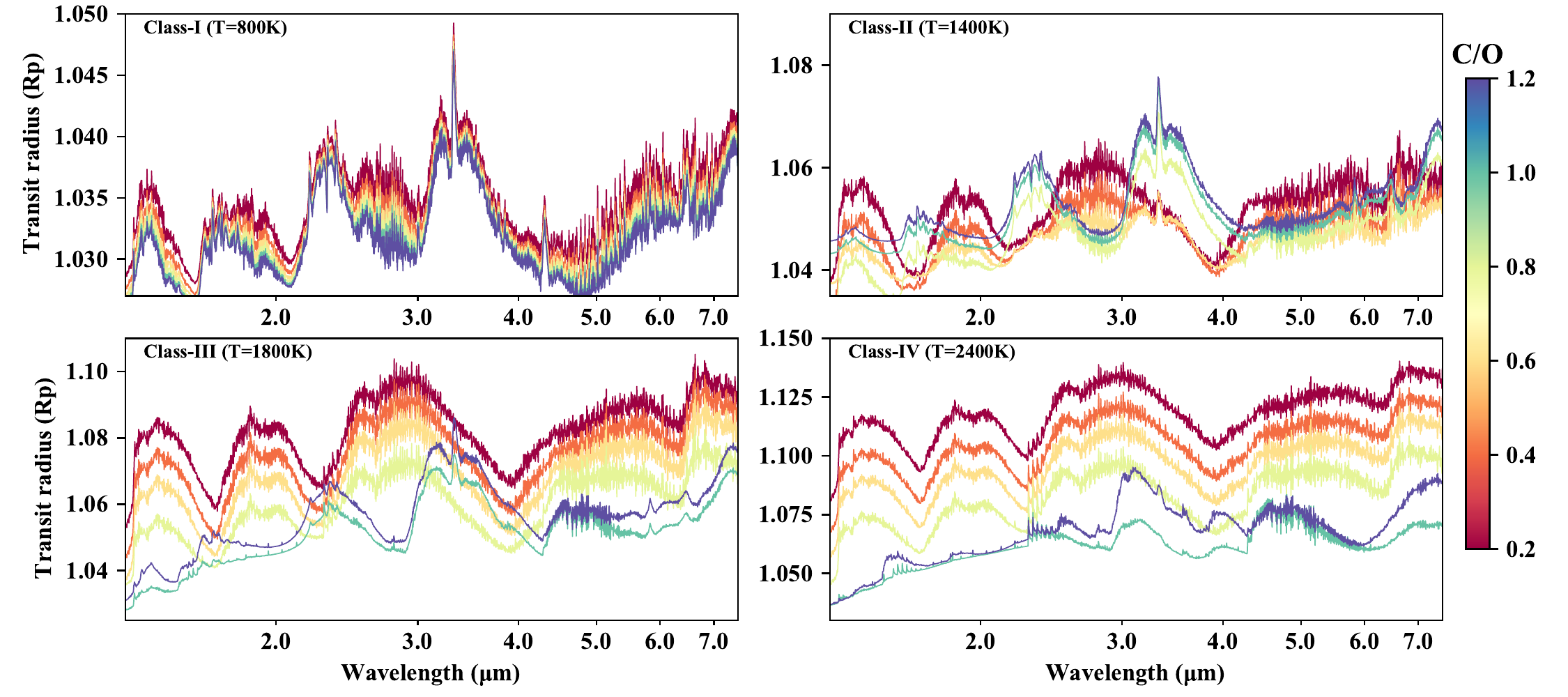}
\caption{Examples of transmission spectra of cloudy models at $f_{\rm sed}\sim$3.0 and [Fe/H]=-1.0.} \label{fig:examplesf300fehn1}
\end{figure*}

Other atmospheric parameters, such as the surface gravity and metallicity, affect the atmospheric response to the variation of sedimentation factor as well. For instance, Figures~\ref{fig:examplesf003fehn1} and \ref{fig:examplesf300fehn1} have a lower metallicity ([Fe/H]=-1.0) in comparison to the examples given in Figures~\ref{fig:examplesf003} and \ref{fig:examplesf300} ([Fe/H]=1.0). A striking difference is the presence of methane. Even the hottest cases, T\textsubscript{eff}=2400\;K, show the contribution of methane in their spectra between 3.0 and 4.0\;$\mu$m. Lower metalicity results in higher cooling efficiency of atmosphere at the photospheric levels. Such environment is favorable for the production of methane; consistent with the presence of methane in Figures~\ref{fig:examplesf003fehn1} and \ref{fig:examplesf300fehn1} (see also Figures~\ref{fig:examplesf010fehn1}, \ref{fig:examplesf030fehn1} and \ref{fig:examplesf100fehn1} in the Appendix, to compare cases with other sedimentation factors).

\subsection{Estimation of Transition C/O Ratios and the Methane Valley}\label{sec:ch5COratios}
\begin{figure}[t]
\includegraphics[width=\columnwidth]{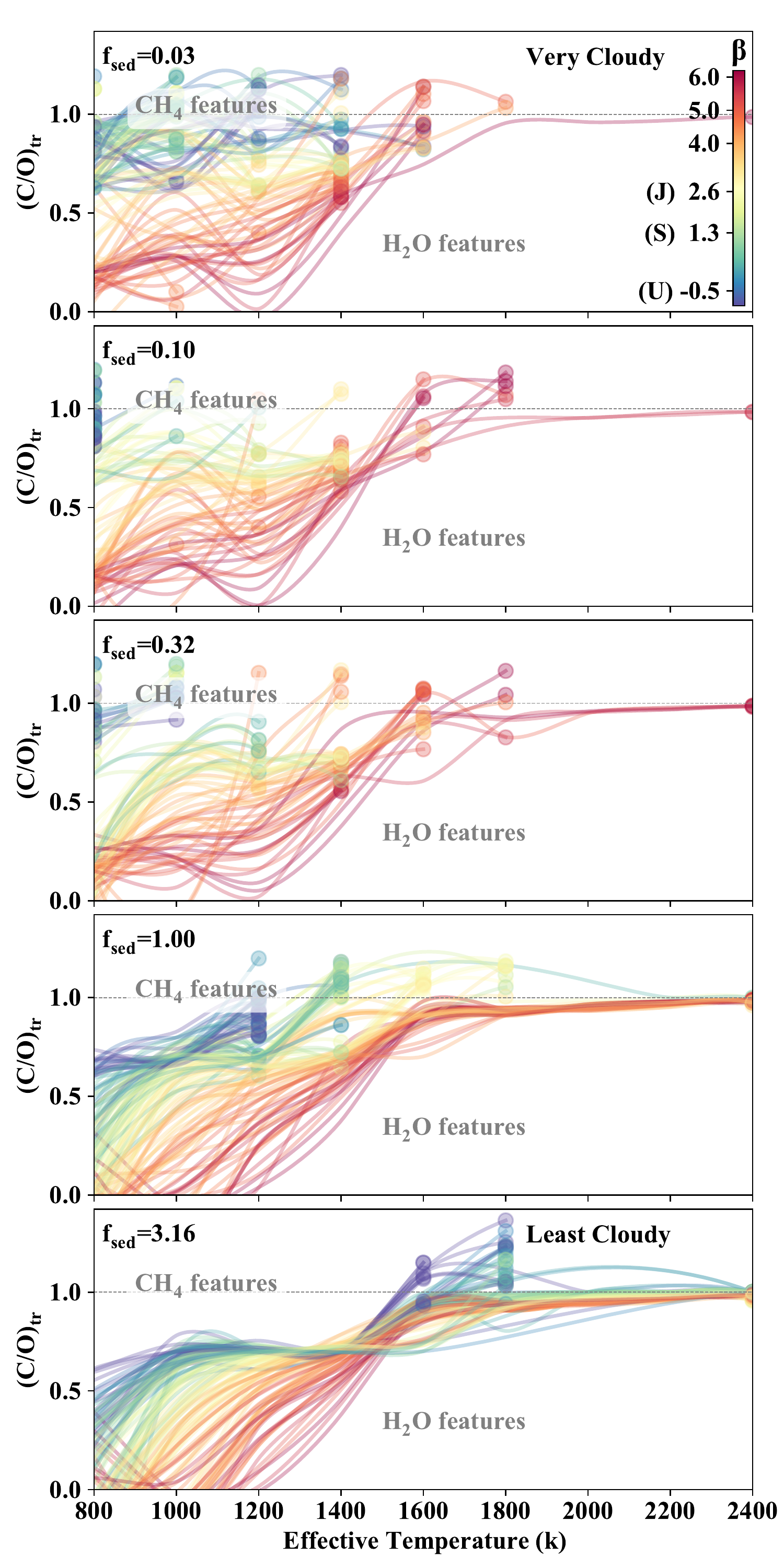}
\caption{Transition carbon-to-oxygen ratios, (C/O)$_{\rm tr}$, for different sedimentation factors, $f_{\rm sed}$. Models above (C/O)$_{\rm tr}$ curves present \ce{CH4} features in their transmission spectra and models below these curves present \ce{H2O} as their dominant spectral feature. The curves are color-coded by the $\beta$-factor of the models ($\beta$=log(g)-1.7$\times$[Fe/H]). The $\beta$-factor of Jupiter (J), Saturn (S), and Uranus (U) are shown for comparison. Circle symbols mark the highest temperature at which \ce{CH4} features is present in the transmission spectra for that given $\beta$-factor and $f_{\rm sed}$. See Section~\ref{sec:ch5COratios} for a discussion.} \label{fig:ch5COtr}
\end{figure}

Our results indicate that clouds can drastically limit the parameter space at which methane is expected to be present in the transmission spectra of exoplanets. The transition C/O ratios at which methane features become the dominant spectral features are estimated by using the spectral decomposition technique. 
Figure~\ref{fig:ch5COtr} shows the corresponding parameter space separated for different sedimentation factors. The transition C/O ratios are color-coded by the $\beta$-factor. $\beta$-factors of Jupiter ($\sim$2.6), Saturn ($\sim$1.3), and Uranus ($\sim$-0.5) are also marked for comparison.

We start by interpreting the results of $f_{\rm sed}$=3.16 models (the bottom panel in Figure~\ref{fig:ch5COtr}). Models below 1650\;K (mostly Class-II, but also Class-I) show transition C/O ratios significantly less than unity. Under these conditions, the Methane Valley is a persisting domain above C/O$\sim$0.7 where methane features are expected to be present in the planetary spectra regardless of their metallicity or surface gravity. (C/O)$_{\rm tr}$ ratio increases by temperature due to partial evaporation of condensates and the release of sequestered oxygen. At any given temperature in this regime, a lower metallicity favors cooling the photosphere and the photosphere moves to larger pressures \citep[e.g. see Figure 11 in ][]{molliere_model_2015}. This increases the amount of methane at the photospheric levels, which in turn lowers the (C/O)$_{\rm tr}$ ratios at higher $\beta$-factors. Above 1650\;K, (C/O)$_{\rm tr}$ ratios remain around the unity for models with higher $\beta$-factors. The lower $\beta$-factors ($\lesssim$2.0), however, have terminated ends on their (C/O)$_{\rm tr}$ ratios (circles at the right-end of (C/O)$_{\rm tr}$ ratio lines). This represents the lack of methane in the transmission spectra of all hotter models with the same $\beta$-factor. This is due to the methane depletion and obscuration by other opacity sources as discussed. While the termination of (C/O)$_{\rm tr}$ ratios at lower $\beta$-factors limits the Methane Valley to colder planets, the results of models with $f_{\rm sed}$=3.16 resemble the results of cloud-free atmospheres in general, as expected \citep[See Section 4 in][]{molaverdikhani_toward_2019}.

Models with $f_{\rm sed}$=1.0 start to differentiate from the cloud-free models and the Methane Valley shrinks for low $\beta$-factor models. Under these conditions, the observation of methane on planets with similar $\beta$-factor to that of Saturn (and Uranus/Neptune) would be limited to temperatures below 1400\;K (and $<$1200\;K). The Methane Valley does not change much for planets with Jovian-like $\beta$-factors as its $\beta$-factor is not very low.

At lower sedimentation factors, $f_{\rm sed}<$1.0, the Methane Valley would vanish for planets with Uranus- or Neptune-like $\beta$-factors. We should note that a common implementation of the \citet{ackerman_precipitating_2001} model (as it is also the case in our grid of cloudy models) is using the same $f_{\rm sed}$ for all species. This could result in an over-estimation of cloud opacities at colder planets. Inclusion of species-dependent $f_{\rm sed}$ may remove some of the condensates from the photosphere to make room for the reappearance of methane in the transmission spectra of colder planets. Nevertheless, our systematic survey provides a framework to investigate the effect of species-dependent rain-out on the observability of methane (and other species) on exoplanets in the future.

\revision{Note that the curves shown in Figure~\ref{fig:ch5COtr} exhibit some oscillations with T\textsubscript{eff}, which manifest themselves as local extrema. While these local extrema at the grid points are caused by highly non-linear cloud feedback, the local extrema in between grid points are artifacts caused by the plotting algorithm to make the curves smooth.}

\section{Spitzer's Color-diagrams}\label{sec:ch5color}

\begin{figure*}
\includegraphics[width=\textwidth]{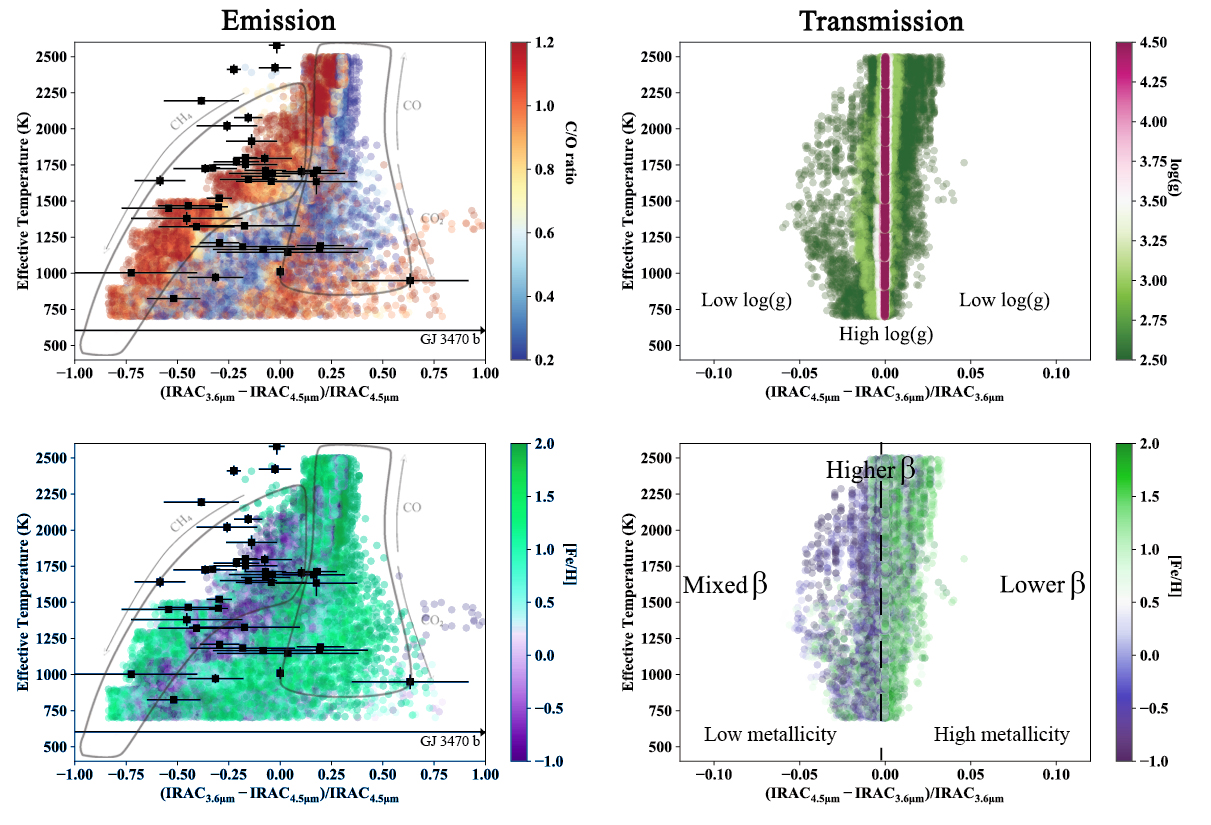}
\caption{Spitzer's synthetic color-diagrams of cloudy irradiated planets in emission (left panels) and transmission (right panels). Emission diagram is compared with the observations (black squares). Cloud-free \ce{CH4} and CO/\ce{CO2} populations are shown by gray lines for reference.
The shape of these populations in cloudy models have shifted: Hotter models are less scattered due to methane depletion and HCN production, and colder planets (with effective temperature less than $\sim$1200\;K) are skewed toward right, where a higher contribution from IRAC channel 1 at 3.6$\mu$m influences the color. The cloudy transmission diagram remains similar to its cloud-free analogue. See Section~\ref{sec:ch5color} for details.} \label{fig:ch5colormap}
\end{figure*}

We construct synthetic Spitzer IRAC color-diagrams by extracting the normalized ``colors'' of our cloudy grid of atmospheric models. These colors are based on IRAC's photometric channels 1 and 2, which are centered at 3.6 and 4.5\;$\mu$m, respectively, and for the transmission diagrams are calculated as below:
\begin{equation}
    \label{eq:IRAC_trans} 
R_{\rm tr}=({\rm IRAC}_{4.5\;\mu {\rm m}}-{\rm IRAC}_{3.6\;\mu {\rm m}})/{\rm IRAC}_{3.6\;\mu {\rm m}}
\end{equation}
and it is the other way around for the emission diagrams:
\begin{equation}
\label{eq:IRAC_emission}
    R_{\rm em}=({\rm IRAC}_{3.6\;\mu {\rm m}}-{\rm IRAC}_{4.5\;\mu {\rm m}})/{\rm IRAC}_{4.5\;\mu {\rm m}}
\end{equation}

Channel 1 (3.6\;$\mu$m) is mostly suitable to study \ce{CH4}/\ce{H2O} spectral features and channel 2 (4.5\;$\mu$m) is more sensitive to CO/\ce{CO2} features. HCN contributes in both channels similarly, with slightly higher contribution in channel\;1. Clouds are expected to change the color of planets significantly \citep[e.g.][]{parmentier_transitions_2016}. They could have either gray or non-gray contributions in these channels, depending on their properties, such as their particle size distribution. Therefore, the presence of populations on the color-diagram could reveal dominant processes at the photosphere of planetary atmospheres from a statistical point of view.

Figure~\ref{fig:ch5colormap} shows the constructed color-diagram of the cloudy atmospheres, both in emission (left panels) and transmission (right panels). In the emission diagrams, two main populations from cloud-free simulations are shown by gray lines: \ce{CH4} (left) and CO/\ce{CO2} (right) dominated populations. The overall shape of the cloudy populations remain similar to that of cloud-free and disequilibrium atmospheres \citep{molaverdikhani_fingerprints_2019} but with some differences. Paucity of methane at hotter planets and its replacement with HCN (mostly in models with temperatures above 1500\;K) pushes the originally populated regions with methane-dominated cloud-free atmospheres to the right. This is because HCN contributes more evenly in IRAC channels than \ce{CH4}; see Figure~\ref{fig:ch5spectemp}. On the other hand, obscuration of CO by HCN results in a color-shift of the hot population with low C/O toward zero from the right-side of the map. In addition to the contribution of HCN instead of CO and \ce{CH4}, gray/semi-gray cloud opacities also cause less contrast between the two channels, which makes the models to populate color values around zero.

Including the available IRAC's photometric observations in the emission maps, i.e. black squares with error bars in Figure~\ref{fig:ch5colormap} left panels, reveals a general agreement between our cloudy models and the observations. However, Class-IV exoplanets exhibit systematically redder colors (negative color values). One possibility is neglecting the effect of 3D geometry in this work. The colder night-side environment of hot exoplanets provides a suitable environment for chemistry that are relevant to the colder regimes, including the formation methane. Such products can be transported to the morning side to contribute in the spectra of these planets \citep[e.g.][]{helling_understanding_2019,molaverdikhani_understanding_2020}. Other biases due to the neglecting of 3D geometry of exoplanet have been also reported \citep{caldas_effects_2019,taylor_Understanding_2020,pluriel_strong_2020}, which emphasises the importance of this effect, in particular, on hotter exoplanets.

In the cloud-free models, the population of colder planets (particularly for $\lesssim$1200\;K) are mostly bounded to the \ce{CH4}-population (gray area at the left with significantly negative color values). Only limited number of these planets with low C/O ratios show higher color values than that of \ce{CH4}-population. This is mainly because of the persisting presence of \ce{CH4} in cold cloud-free models. As shown by \citet{molaverdikhani_fingerprints_2019}, introducing disequilibrium processes does not change this picture, but in this work we show that clouds affect this population remarkably. In the cloudy models, the population is scattered toward right, where a higher contribution from IRAC channel\;1 at 3.6$\mu$m influences the color. The reason is the presence of clouds; through methane depletion and obscuration of molecular features. In addition, high-metallicity planets favor the formation of \ce{CO2}, which in turn smears the population to bluer colors even more. Studying this parameter space could therefore help determining the presence and statistics of clouds on exoplanets as their colors at these wavelengths would differ from their cloud-free counterparts considerably.

GJ\;3470\;b is a sub-Neptune with an equilibrium temperature around 600\;K; solid line in the emission maps marks its location. \citet{benneke_sub-neptune_2019} performed photometric measurements of the thermal emission of this planet with IRAC channel 1 ($F_p/F_*$=115$\pm$27~ppm) and 2 ($F_p/F_*$=3$\pm$22~ppm). They argued that the high contrast in these two channels and a very low flux in channel\;2 is in contrast with solar abundance models and a low methane abundance is in favor. They also noted that photochemistry at upper layers cannot explain this contrast and possible explanations are 1) substantial interior heating, 2) photochemical destruction of \ce{CH4} at deeper levels, or 3) a low C/O ratio as a result of the planet formation process. \citet{molaverdikhani_fingerprints_2019} showed a low C/O ratio could indeed explain the emission color of some of the low temperature cloud-free planets that are skewed to the right of the map. However, that alone would be insufficient to explain the observations of GJ\;3470\;b. While our cloudy grid is not extended to 600\;K, it is apparent that at temperatures lower than 800\;K the contribution from the clouds would continue to skew the emission map to the right. Therefore, the role of clouds could be a more likely explanation for the photometric observation of GJ\;3470\;b as opposed of a lower C/O (unless its atmospheric C/O ratio is extremely low). We should, however, note that an internal heat could be still a plausible explanation in this case, which requires further investigations.

Figure~\ref{fig:ch5colormap} lower-left panel shows the same emission color-diagram but color-coded by metallicity instead of C/O ratio. The scattered population of colder planets ($\lesssim$1200\;K) seems to be associated with higher metallicities in general. This is consistent with the depletion of \ce{CH4} at higher metallicities as also summarized in Figure~\ref{fig:ch5COtr}. As noted, higher metallicity is also in favor of a higher \ce{CO2}/\ce{H2O}, which enhances the shift toward more positive colors. It is worth noting that such ``hidden metallicity'' cannot be estimated directly from the current cloudy or cloud-free self-consistent retrievals as they do not consider any physicochemical feedback from the clouds during the forward model calculations. Consequently, an obscured spectrum by clouds could be interpreted as a low-metallicity atmosphere through these approaches.

Transmission color-map (right panels of Figure~\ref{fig:ch5colormap}) does not change significantly from that of cloud-free and disequilibrium; i.e. planets with lower surface gravity would have a more disperse colors and high log(g) would make the spectral features less pronounced and clustered around zero. Note that the grid of cloudy models is extended from log(g) of 2.5 to 4.5; hence a less scattered models are noticeable relative to the previous models with a log(g) of 2.0 to 5.0 in our cloud-free investigation \citep{molaverdikhani_toward_2019}. A final remark on the transmission map is that, in general, planets with higher metallicities tend to populate the right side of the map and otherwise for the low metallicity atmospheres. Therefore, a combination of low metallicity and low surface gravity (i.e. low $\beta$-factor) could potentially explain some of the observations at the far-right region of the transmission map; see \citet{baxter_comprehensive_2018} for some examples.

\section{Conclusion}\label{sec:ch5conclusion}
We calculated a grid of 37,800 self-consistent cloudy atmospheric models and investigated the role of clouds in the classification of planets based on their transmission spectra. Our quantitative analysis confirms that our proposed classification scheme based on cloud-free models in \citep{molaverdikhani_toward_2019} holds true in the case of cloudy atmospheres, in principal. However, alteration of this classification by the presence of clouds is noticeable. For instance, high-metallicity and low sedimentation factor could affect the the dominant chemical processes at the photospheric levels and change the transmission spectra significantly.

For a given atmosphere, the significance of methane's spectral signature can change in two ways: by changing its abundance or by obscuration due to the presence of other opacities (whether in gas- or solid-phase). We found that in all carbon-poor cold cloudy planets (Class-I) the transmission spectra are void of methane. This is in contrast with our previous findings based on cloud-free models where methane appears in the spectra of Class-I planets regardless of their C/O ratio. We linked the heating of the photosphere to this depletion of methane that is caused by optically thick layers of clouds in the models. The heating is more efficient when the metallicity is high. Such heating partially releases the sequestered oxygen from the condensates; making the transmission spectra of cold carbon-poor exoplanets to be more likely rich in water.

We predict that an \revision{observational} consequence of such photospheric heating is simultaneous depletion of \ce{NH3} and \ce{CH4} from the photosphere. Photochemical reactions could also deplete these species. However, determination of clouds' composition could distinct between photochemically formed soot haze and silicate, alkali sulfates, salts, or any other non-soot clouds.

Hotter planets result in optically thinner clouds at their photospheres, which leads to a cooler photosphere, due to a weaker greenhouse effect. This cooling, however, does not results in significant methane production as the temperatures are still much higher than suitable temperatures for \ce{CH4} formation. HCN and CO, on the other hand, form efficiently both on hotter planets and colder planets with heated photospheres. Hence HCN features are expected to be more prominent in the atmosphere of carbon-rich (high C/O) atmospheres, and in particular on high-metallicity planets. The detection of HCN is observationally favorable as its spectral features contrast \ce{H2O} in the range of 3-5$\mu$m \citep[e.g.][]{macdonald_signatures_2017}. A lack of \ce{CH4} and the presence of HCN may be another indication of cloud formation in the atmosphere of exoplanets.

In general, the detection of methane on exoplanets with $\beta$-factors similar to Jupiter's and Saturn's is more likely than planets with $\beta$-factors similar to Uranus's and Neptune's. In addition, higher sedimentation factors remove the clouds from the photospheres more efficiently. Thus atmospheric properties resemble cloud-free atmospheres. In contrast, lower $f_{\rm sed}$ forms extended clouds that heat the photosphere more, deplete methane and mute its remaining signatures from the transmission spectra. Under these suppressed conditions, observation of methane at NIR bands is a challenging task. We therefore propose observing its fundamental band at 3.3\;$\mu$m as has been also proposed by others \citep[e.g.][]{noll_onset_2000,kawashima_detectable_2019}. Nevertheless, low metallicity planets, e.g. HAT-P-11b \citep[][]{chachan_hubble_2019}, are in favour of methane detection for the future observations.

All together, higher surface gravity, lower metallicity, and higher sedimentation factor seems to \revision{provide} the most suitable condition to find methane on irradiated exoplanets. \revision{While surface gravity and metallicity are properties of exoplanets, $f_{sed}$ acts as a proxy for the vertical extent of the clouds and is a model parameter. The effects of $f_{sed}$ can be altered by changing other input parameters of the Ackerman and Marley model, which we do not investigate in this work but should be taken into account when interpreting the results of this grid. Two examples of such atmospheric parameters are the particle size distribution and vertical mixing strength.}

\revision{Nevertheless,} assuming an anti-correlation between the metallicity and the mass of exoplanets \citep[e.g.][]{wakeford_complete_2017}, more massive planets are then in favor of methane detection. Combining this with higher surface gravity condition results in \textbf{smaller but massive planets} to be the favorable targets. Moreover, slow rotating planets are more likely to have less rigorous atmospheric mixing than fast rotating planets \citep[e.g.][]{zhang_globalmean_2018}. This favors higher efficiency in cloud settlement. A slow rotating tidally-locked exoplanet translates to a longer orbital period. Therefore, planets with longer orbital periods are more likely to show methane in their spectra. On the other hand, the hottest planets in the range of Methane Valley's temperature range (i.e. $\sim$1450~K) are more likely to have thinned cloud layers while demonstrating methane in their spectra. Thus planets with \textbf{temperatures around 1450~K} with longest orbital periods could be favorable candidates for methane detection. This also translates to the planets with T$\sim$1450~K around the coldest host starts, e.g. M-dwarfs, as their orbital periods are the longest. These are only general guidelines and further observations are needed to proof these predictions.

\begin{figure*}
\includegraphics[width=\textwidth]{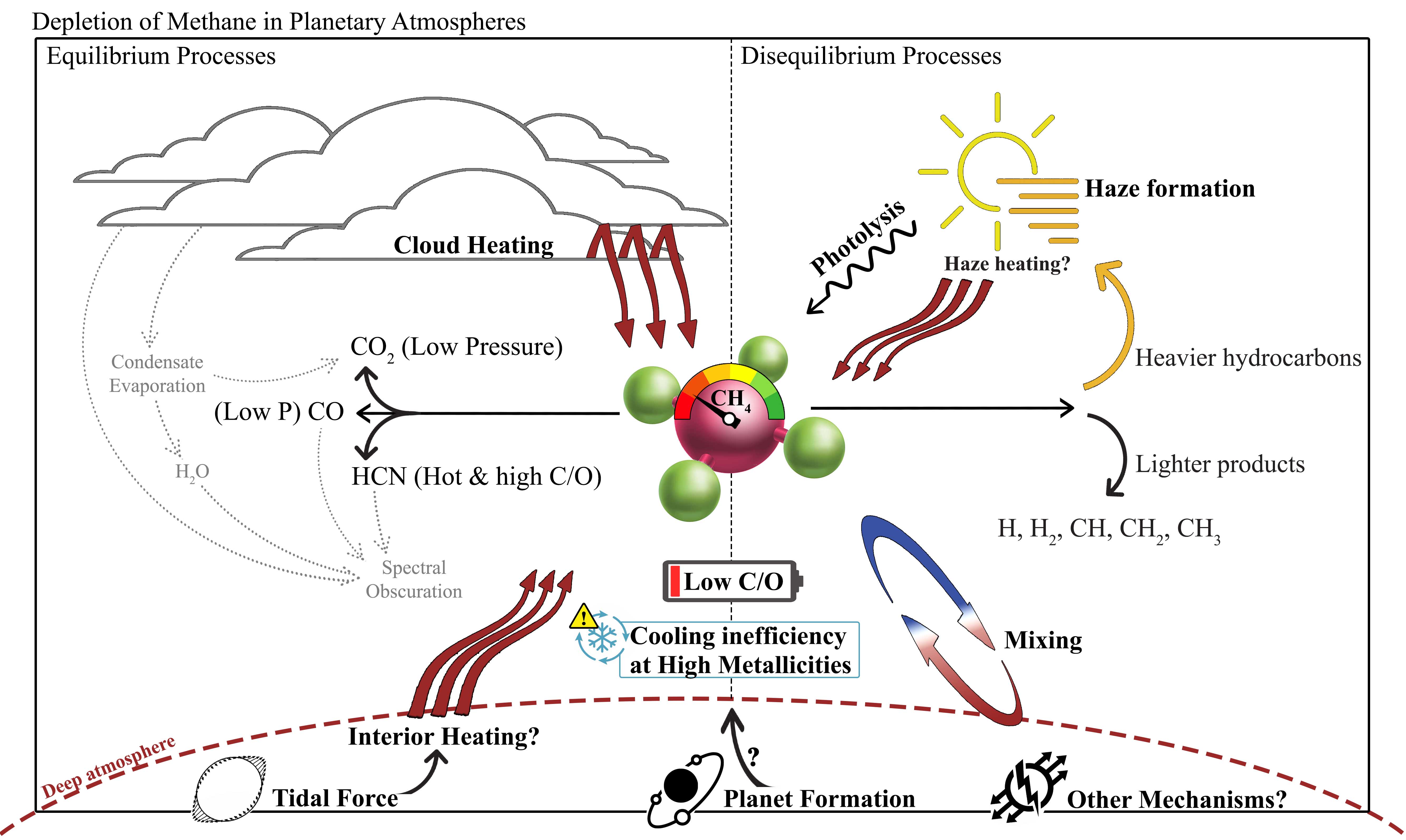}
\caption{A summary of major atmospheric processes that cause methane depletion. Some of the icons are adapted from Colourbox.com, 123RF.com, depositphotos.com, and gettyimages.com.} \label{fig:ch5final}
\end{figure*}

A summary of major atmospheric processes that cause methane depletion are also illustrated in Figure~\ref{fig:ch5final}. This includes heating by clouds and hazes, internal heating by tidal force or remnant heating from the formation or other mechanisms, lower C/O ratio or high metallicity through planetary formation, atmospheric mixing, photolysis, etc.

We also investigated the shape of populations in the emission and transmission color-diagrams due to the cloud formation. At higher temperatures the populations tend to cluster around lower contrast values relative to the cloud-free and disequilibrium emission maps. We found that this is mainly due to the conversion of \ce{CH4} and CO into HCN. Since HCN has more even contribution in the IRAC channels, the scattered population of hot planets becomes more clustered. At low temperatures, however, the colors are more dispersed and skewed toward right, where the depletion of methane and higher \ce{CO2} contributions in high-metallicity planets makes the normalized color to become more positive (bluer in a non-normalized sense). This scattered population of cold planets can be used to characterize the formation of clouds on exoplanets through multi-color photometry using facilities that are more accessible than spectrographs.





\section{Acknowledgment} \label{sec:acknowledgment}
T.H. and P.M. acknowledge support from the European Research Council under the Horizon 2020 Framework Program via the ERC Advanced Grant Origins 83 24 28. This research has made use of the NASA Exoplanet Archive, which is operated by the California Institute of Technology, under contract with the National Aeronautics and Space Administration under the Exoplanet Exploration Program.

\appendix

\section{Additional Figures} \label{sec:ch5appfig}

\begin{figure*}
\includegraphics[width=\textwidth]{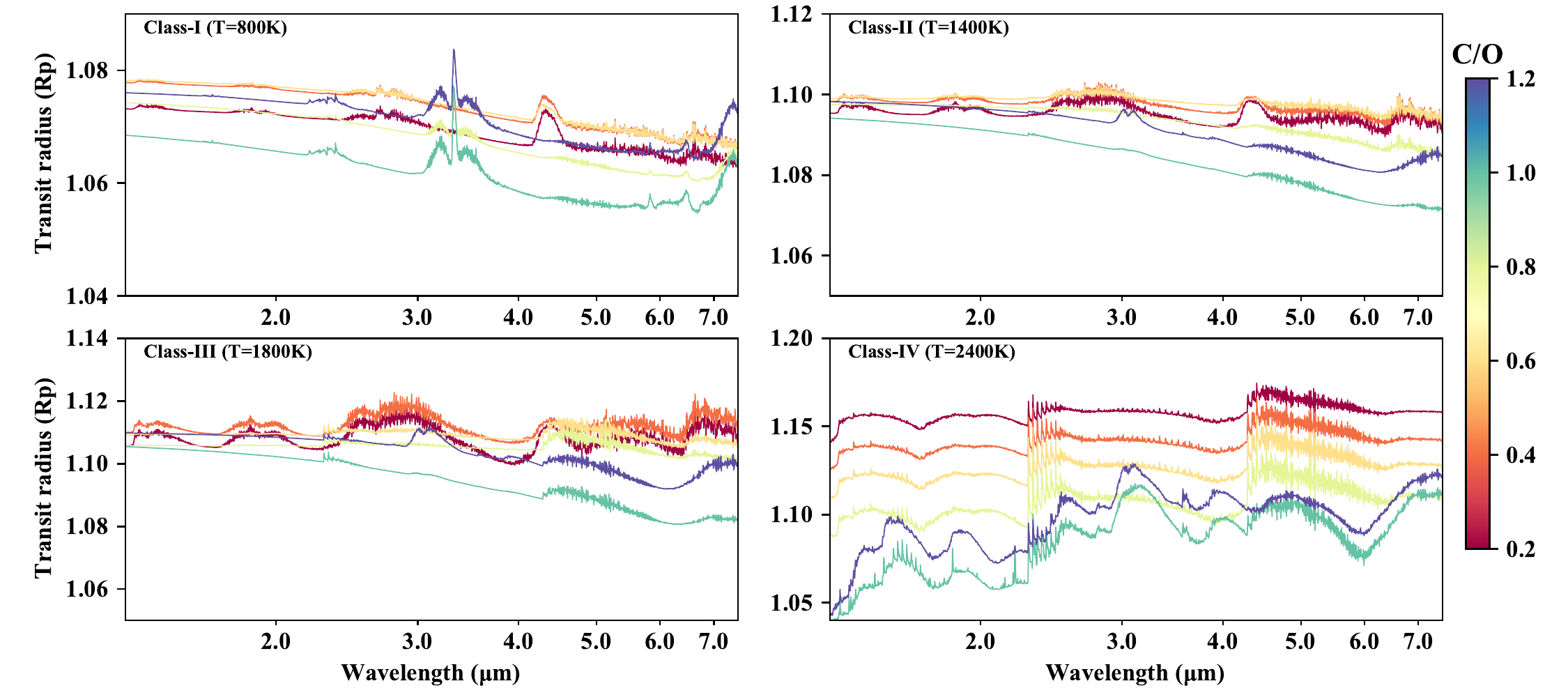}
\caption{Examples of transmission spectra of cloudy models at $f_{\rm sed}\sim$0.1.} \label{fig:examplesf010}
\end{figure*}

\begin{figure*}
\includegraphics[width=\textwidth]{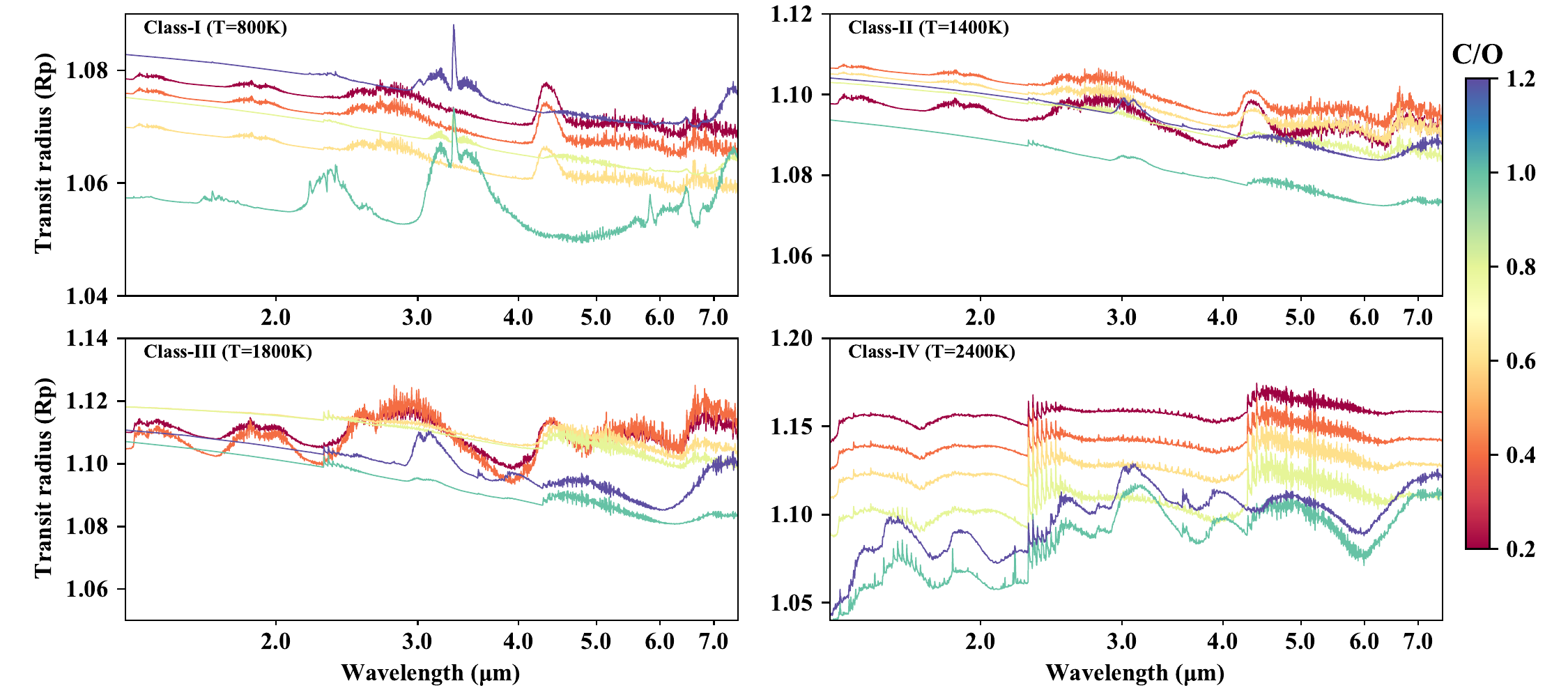}
\caption{Examples of transmission spectra of cloudy models at $f_{\rm sed}\sim$0.3.} \label{fig:examplesf030}
\end{figure*}

\begin{figure*}
\includegraphics[width=\textwidth]{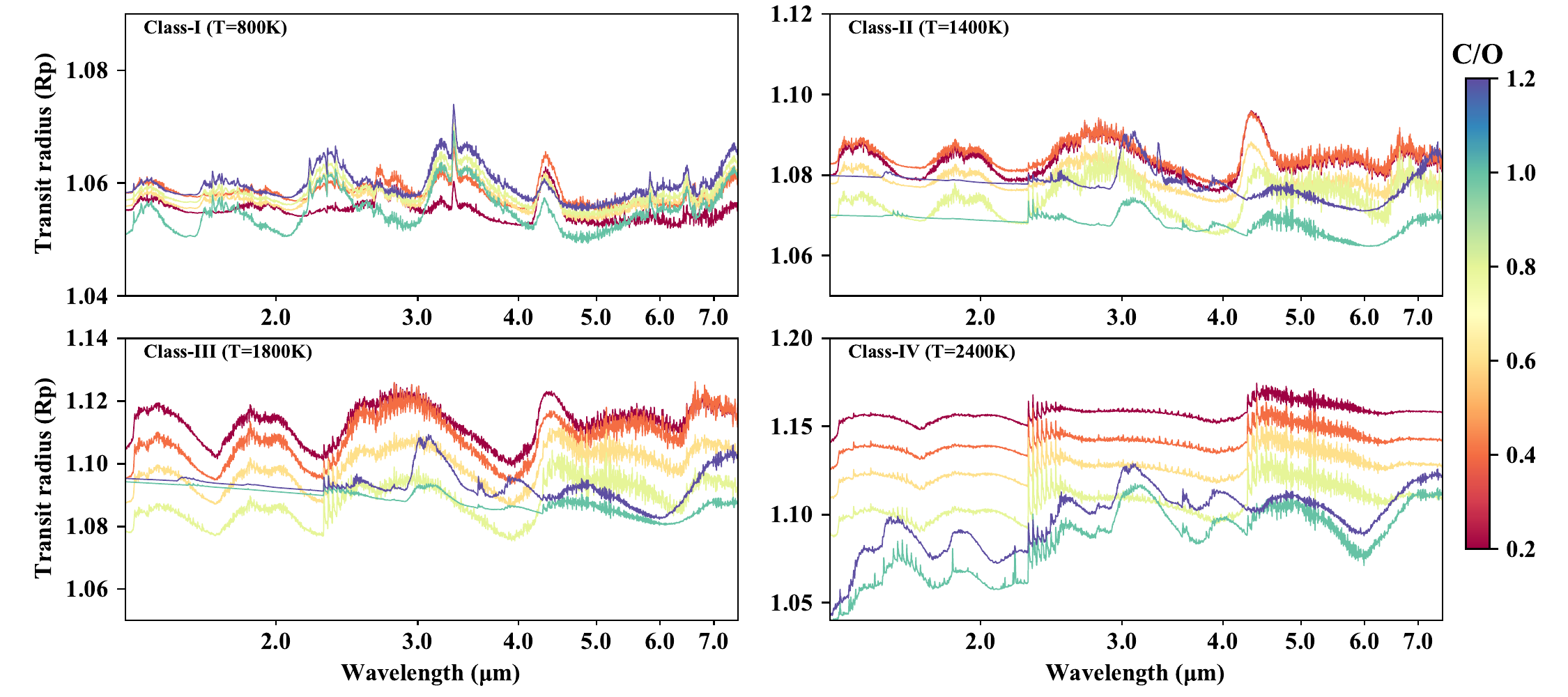}
\caption{Examples of transmission spectra of cloudy models at $f_{\rm sed}\sim$1.0.} \label{fig:examplesf100}
\end{figure*}

\begin{figure*}
\includegraphics[width=\textwidth]{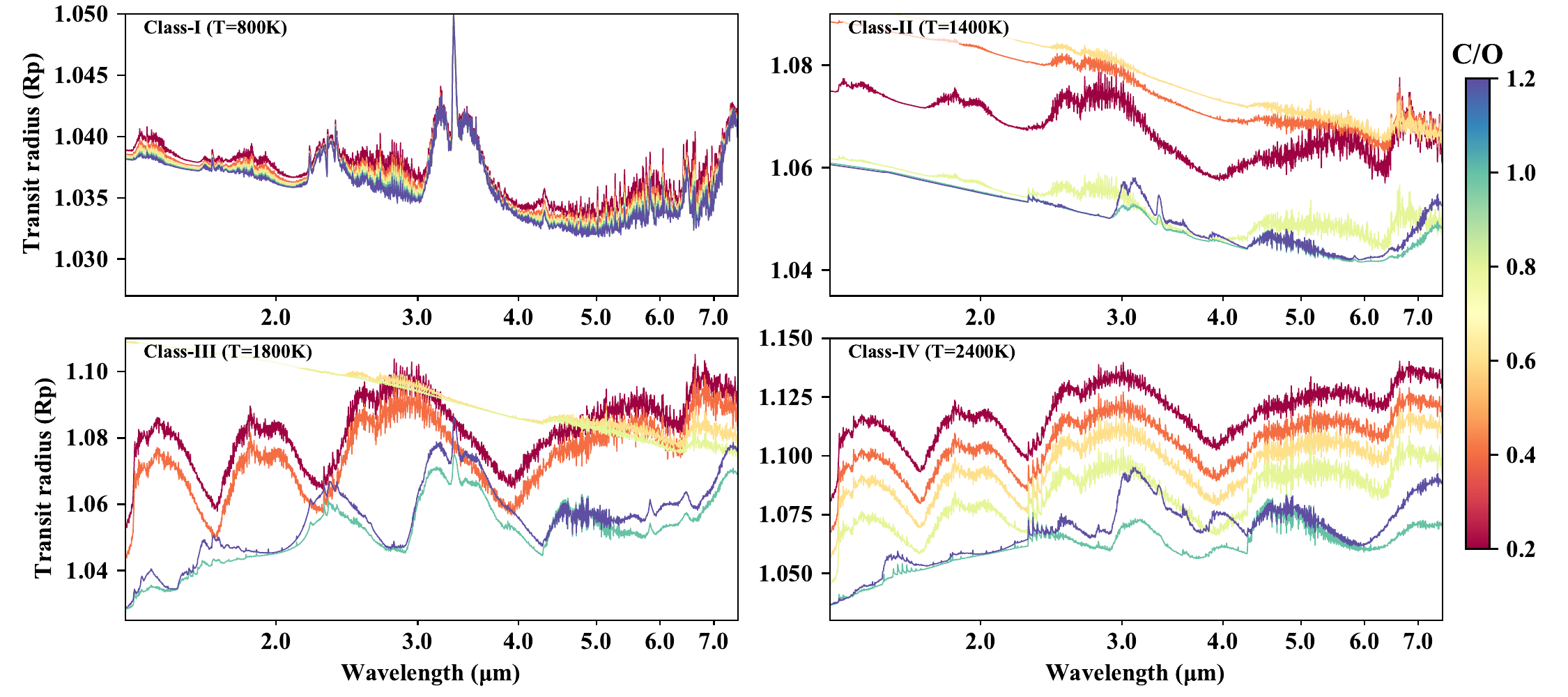}
\caption{Examples of transmission spectra of cloudy models at $f_{\rm sed}\sim$0.1 and [Fe/H]=-1.0.} \label{fig:examplesf010fehn1}
\end{figure*}

\begin{figure*}
\includegraphics[width=\textwidth]{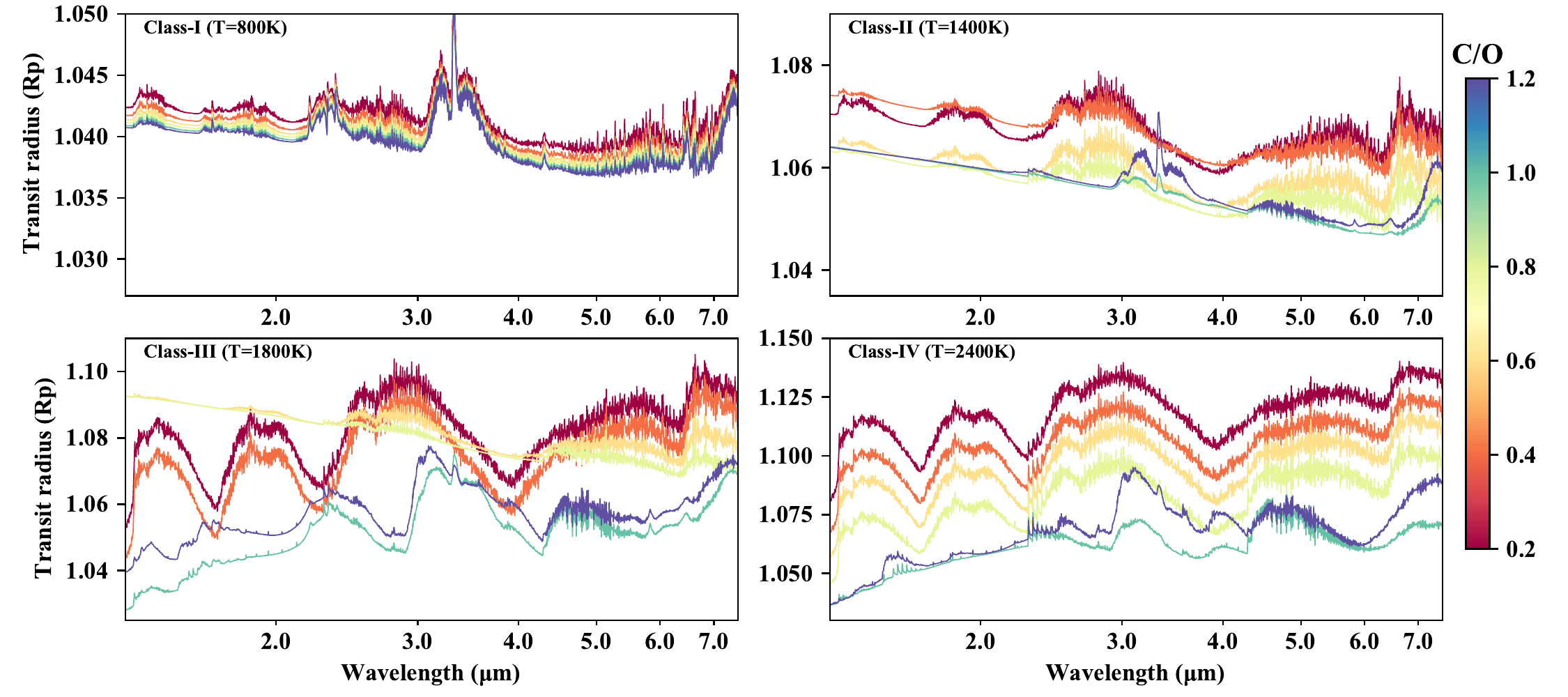}
\caption{Examples of transmission spectra of cloudy models at $f_{\rm sed}\sim$0.3 and [Fe/H]=-1.0.} \label{fig:examplesf030fehn1}
\end{figure*}

\begin{figure*}
\includegraphics[width=\textwidth]{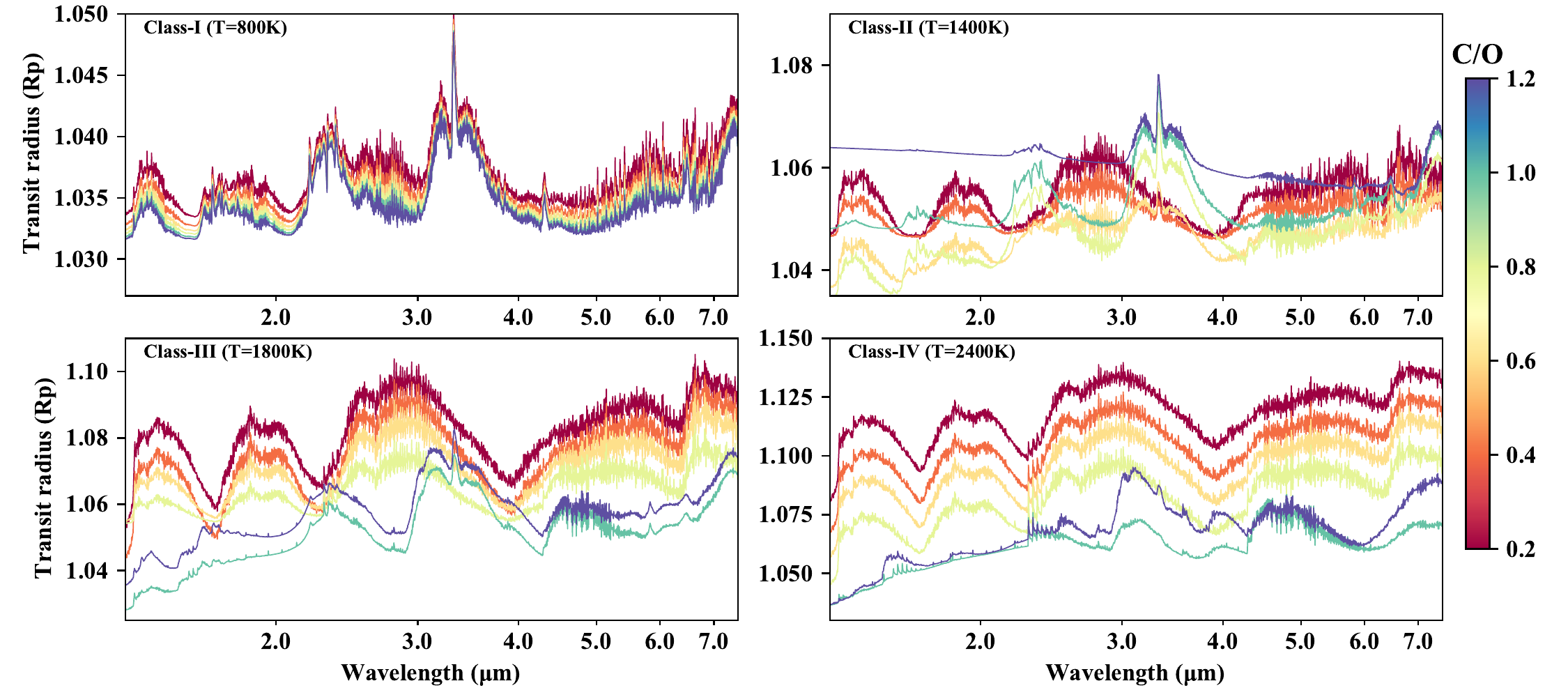}
\caption{Examples of transmission spectra of cloudy models at $f_{\rm sed}\sim$1.0 and [Fe/H]=-1.0.} \label{fig:examplesf100fehn1}
\end{figure*}

Figures~\ref{fig:examplesf010}, \ref{fig:examplesf030}, \ref{fig:examplesf100}, \ref{fig:examplesf010fehn1}, \ref{fig:examplesf030fehn1} and \ref{fig:examplesf100fehn1} are shown as the examples of how sedimentation factor affects the transmission spectra of exoplanets at different metallicities and for different classes of planets.

\bibliographystyle{aasjournal}
\bibliography{paper}{}



\end{document}